%% file: main.tex
\documentclass[10pt]{iopart}

\usepackage{cite}
\usepackage{booktabs}
\usepackage[breaklinks=true,
            colorlinks=true,
            linkcolor=blue,
            urlcolor=blue,
            citecolor=blue,
            bookmarksopen=true,
            bookmarksopenlevel=2,
            bookmarksnumbered=true,
            pdfauthor=Maximilian Dreisbach,
            pdfstartview=FitH]
            {hyperref}
\usepackage{tikz,pgfplots}
\usetikzlibrary{spy,backgrounds}
\pgfplotsset{compat=1.9}
\usepackage[mode=buildnew]{standalone}
\usepackage{comment}
\usepackage{appendix}
\usepackage{xcolor}
\usepackage{amssymb}
\usepackage{subcaption}
\usepackage{textgreek}

\usepackage{stfloats}

\definecolor{KITgreen}{RGB}{0 150 130}
\definecolor{KITblue}{RGB}{70 100 170}
\definecolor{KITmaygreen}{RGB}{140 182 60}
\definecolor{KITyellow}{RGB}{252 229 0}
\definecolor{KITorange}{RGB}{223 155 27}
\definecolor{KITbrown}{RGB}{167 130 46}
\definecolor{KITred}{RGB}{162 34 35}
\definecolor{KITpurple}{RGB}{163 16 124}
\definecolor{KITcyan}{RGB}{35 161 224}

\newcommand{\citet}[1]{\emph{et al.}\,\cite{#1}}

\begin{document}

%\title{Adhering droplet interface reconstruction for distortion correction using glare points and deep learning}
%
\title{Interface reconstruction of adhering droplets for distortion correction using glare points and deep learning}
%Spatio-temporal reconstruction of an adhering droplet through deep learning and glare points

\author{Maximilian Dreisbach$^{1,\footnotemark}$, Itzel Hinojos$^{1}$, Jochen Kriegseis$^1$, Alexander Stroh$^{1}$, Sebastian Burgmann$^{2}$}
\footnotetext{Author to whom any correspondence should be addressed}

\address{$^1$ Institute of Fluids Mechanics (ISTM), Karlsruhe Institute of Technology (KIT), Kaiserstraße 10, 76131 Karlsruhe, Germany,
$^2$ Chair of Fluid Mechanics, University of Wuppertal, Gaußstraße 20, 42119 Wuppertal, Germany}

\vspace{10pt}
\begin{indented}
\item[]January 2025
\end{indented}

\begin{abstract}
The flow within adhering droplets subjected to external shear flows has a significant influence on the stability and eventual detachment of the droplets from the surface.
Most commonly, the velocity field inside adhering droplets is measured by means of particle image velocimetry (PIV), which requires a correction step to account for the distortion caused by the refraction of light at the curved gas-liquid interface.
Current methods for distortion correction based on ray tracing are limited to low external flow velocities, for which the deformation of the droplet is insignificant and axisymmetry can be assumed.
However, the ray-tracing method can be extended straightforwardly to arbitrarily deformed droplet shapes if the instantaneous three-dimensional droplet interface can be acquired.
In the present work, a previously introduced method for the image-based reconstruction of gas-liquid interfaces by means of deep learning is adapted to determine the instantaneous interface of adhering droplets in external shear flows.
In this regard, a purposefully developed optical measurement technique based on the shadowgraphy method is employed that encodes additional three-dimensional (3D) information of the interface in the images via glare points from lateral light sources.
On the basis of the images recorded in the experiments, the volumetric shape of the droplet is reconstructed by a neural network that was trained on the spatio-temporal dynamics of the gas-liquid interface from a synthetic dataset obtained by numerical simulation.
The results for experiments with adhering droplets at different velocities of external flow demonstrate that the combination of the learned droplet geometry with the depth encoding through the glare points facilitates a robust and flexible reconstruction.
The proposed method reconstructs the instantaneous three-dimensional interface of adhering droplets at both high resolution and spatial accuracy and thereby enables the distortion correction of PIV measurements at high external flow velocities.

\end{abstract}

\noindent{\footnotesize{ \it Keywords}: Two-phase flow, Adhering droplet, Shear flow, Interface reconstruction, Deep learning}
%\submitto{International Journal of Multiphase Flow}
%\maketitle
\ioptwocol

%%%---------------------------------------------------------------------
%Textbody

\section{Introduction}

%Relevance
The internal flow within droplets or bubbles plays an important role in numerous different multiphase flow phenomena, such as droplet dispersing, adhering droplets subjected to external flows, the evaporation or freezing of droplets, and bubble flow among others.
Particle image velocimetry (PIV) is commonly employed in the investigation of multiphase flows to reveal internal flows.
Karlsson~\citet{Karlsson2019} used PIV to visualize the flow patterns and measure velocities inside freezing droplets. 
The authors found that Marangoni effects, which arise due to temperature differences in and around the droplet during freezing, induce a flow inside the droplet.
Kinoshita~\citet{Kinoshita2007} revealed the complex three-dimensional circulation patterns within a moving droplet in a microchannel using micro-PIV.
Later, Duxenneuner~\citet{Duxenneuner2014} investigated the flow inside and around a droplet during liquid-liquid dispersion using micro-PIV and revealed the existence of a flow inside the droplet in main flow direction as well as a vortex flow around the droplet.
By means of PIV measurements Minor~\citet{Minor2009} and later Burgmann~\citet{Burgmann2021} have found that there is a complex interaction between the internal and external flow of adhering droplets, that is dependent on interface tension and the oscillation of both phases.

\begin{figure}[htbp!]
\centering
    \includegraphics[width=\linewidth]{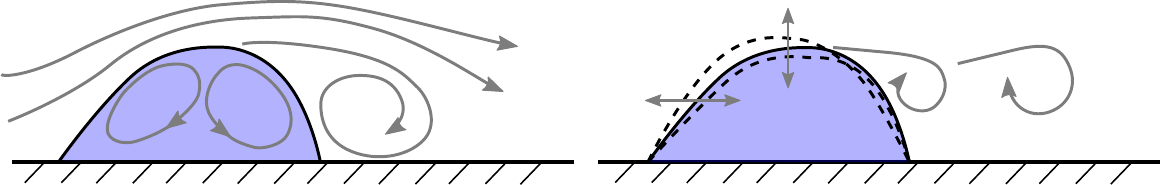}
    \caption{Visualisation of an adhering droplet subjected to an external shear flow; time-averaged flow topology inside the droplet and in the external flow (left), droplet and flow oscillation (right). Figure adapted from Burgmann~\citet{Burgmann2021}.}
    \label{fig:droplet_flow}
\end{figure}

%Adhering droplets in shear flows
Adhering droplets in external flows occur in many technical applications, for instance in cleaning and drying processes \cite{Thoreau2006,Seevaratnam2010}, oil recovery \cite{Thompson1994,Schleizer1999,Gupta2008,Madani2014}, heat exchangers \cite{Korte2001,Kandlikar2002,Wang2022}, airfoil icing prevention \cite{Karlsson2019} and in fuel cells, where a removal of the droplets is crucial for an efficient operation \cite{Theodorakakos2006,Kumbur2006,Esposito2010,Burgmann2013}.
The flow around the droplet eventually leads to a detachment of the droplet from the solid surface at critical conditions, which occur when the external forces acting on the droplet become greater than the adhesion of the liquid droplet to the surface \cite{Fan2011,Fu2014,Barwari2019,Burgmann2021}.
Several investigations have revealed an intricate interaction of the droplet with the surrounding flow.
The droplet deforms \cite{Gupta2008,Seevaratnam2010,Barwari2018} and additionally oscillates, when subjected to external shear flows \cite{Lin2006,Burgmann2018a}. 
It was shown that the contact angle hysteresis of the substrate and the droplet volume have an influence on the critical velocity for droplet detachment \cite{DussanV1987,Kumbur2006,Barwari2019}.

Previous investigations based on PIV-mea\-su\-re\-ments revealed that the internal flow within the droplet is relevant to understanding the mechanism of droplet detachment. %\cite{Minor2009,Burgmann2018a,Burgmann2021}
It was found that the internal flow follows the main flow direction of the external flow \cite{Duxenneuner2009} and exhibits a clockwise rotational flow pattern \cite{Minor2009}.
Minor~\citet{Minor2009} deduced that the secondary flow inside a droplet is induced by shear forces on the interface resulting from the external flow.
Burgmann~\citet{Burgmann2018a,Burgmann2018b} confirmed the findings of a clockwise vortex at low external flow velocities but also found an additional counterclockwise rotation at higher flow velocity that becomes dominant when approaching the critical velocity.
Furthermore, the authors found an oscillation of the droplet with increasing amplitude towards higher external flow velocities.
%The frequencies of this oscillation match the resonant frequency of the droplet and its harmonic frequencies. 
Burgmann~\citet{Burgmann2021} proposed that the droplet detachment is associated with a self-excitation process resulting from the oscillation of the gas-liquid interface, as well as the oscillation of the inner flow structure and the flow field around the droplet.
The authors demonstrated that an emerging backflow region at higher external flow velocities induces the change in the flow pattern within the droplet and that the wake flow oscillates with the same frequency as the droplet.
Later, Burgmann~\citet{Burgmann2022} found that the coupling of the internal and external flow leads to increased flow separation in comparison to rigid bodies and, consequently, a higher pressure force driving droplet detachment.
Bilsing~\citet{Bilsing2024b} measured the internal three-dimensional (3D) flow field by means of microscopic particle tracking velocimetry (PTV) through a transparent substrate from below and thereby revealed that the internal flow topology consists of multiple 3D vortical structures that change drastically depending on the external flow velocity.
In experiments with rigid droplet models, Zhang~\citet{Zhang2021} found that the external flow topology is significantly influenced by the shape of the adhering droplet. 
Consequently, a better understanding of the droplet deformation is required for any future investigation of the external flow.
A visualization of the flow phenomena inside and around the droplet is shown in Figure~\ref{fig:droplet_flow}.
The mutual dependence of wake flow, contour oscillation, and internal flow structure influences the stability of the droplet and eventually leads to the detachment of the droplet at critical conditions.
While there has been significant scientific progress in recent years, the exact mechanism of droplet detachment still remains to be fully understood.
Therefore, further experiments are required to obtain a better analytical model of droplet stability that accounts for aeroelastic effects and, consequently, allows for a more accurate prediction of the onset of droplet motion.

%Distortion correction
Despite the extensive use of PIV and micro-PIV measurements for the investigation of multiphase flows it is well known that refraction at the curved gas-liquid interfaces causes a distortion of the measured velocity fields \cite{Karlsson2019, Kang2004, Minor2007}.
The gas-liquid interface acts as a convex lens that refracts the light that is emitted from seeded particles within the droplet, which are illuminated by a light sheet \cite{Kang2004}.
The refraction at different incident angles along the curved interface leads to a spatially inhomogeneous distortion, characterized by a large magnification in the center and an increasing warping towards the contour of the droplet.
As a consequence, the topology of the internal flow is represented falsely in uncorrected PIV measurements.
While a measurement from below avoids the distortion by the curved interface of the droplet \cite{Bilsing2024b}, most substrates of interest for the technical application are optically opaque and, therefore, inaccessible by this method.
Kang~\citet{Kang2004} and later Minor~\citet{Minor2007} developed a method for distortion correction that employs ray tracing to derive a mapping function between the image plane and the object plane, i.e. the illuminated particles in the light sheet.
This mapping function can be applied to the PIV images or the measured velocity fields directly to reverse the distortion and thereby significantly increase the accuracy of the measurement.
% To obtain an accurate mapping function for each image pair in the experiments, the instantaneous three-dimensional shape of the gas-liquid interface has to be determined \cite{Burgmann2022}.

In the method introduced by Kang~\citet{Kang2004} the contour of the droplet is detected from shadowgraph images and, subsequently, the three-dimensional droplet interface is obtained under the assumption of axisymmetry.
To derive the mapping function between the image plane and the object plane, normally incident light rays on the image plane are traced to the droplet's interface, where the refraction is calculated by Snell's law.
%Kang~\citet{Kang2004} detect the contour of the droplet from shadowgraph images and fit a cosine series using least-squares in order to obtain a continuous analytical representation of the contour.
%The authors assume axisymmetry of the three-dimensional droplet interface and consider geometric optics for the propagation of light.
%The normally incident light ray to any point on the image plane is traced to the interface of the droplet, at which the change in direction through refraction is calculated by Snell's law.
%Subsequently, the refracted ray is traced further to its intersection with the object plane.
%This process is repeated for all points in the image plane to derive the mapping function between the image plane and the object plane.
Through this method up to $80\%$ of the internal flow field can be retrieved successfully, depending on the relative refractive indices of the considered fluids.
Close to the contour, the retrieval of light from within the droplet is inhibited by total internal reflection and, consequently, the velocity field cannot be measured in these regions.
%The velocity field obtained from PIV measurements inside an evaporating droplet after distortion correction can be seen in Figure~\ref{fig:distortion_correction} in comparison to the uncorrected velocity field.
Minor~\citet{Minor2009} demonstrated the applicability of the ray-tracing correction for micro-PIV measurements of the internal velocity field within an adhering droplet subjected to an external flow.
Due to drag forces acting on the gas-liquid interface, the shape of the droplet deviates significantly from the assumed axisymmetrical shape.
The authors stated that adaptive droplet shape approximations are required to reach a higher degree of accuracy for the deformed droplet shape.
% at high external flow velocities.
Burgmann~\citet{Burgmann2022} further underlined that the oscillating contour of adhering droplets at higher external flow velocities hinders an application of the ray-tracing method.
A possible solution for the correction of the dynamic aberration by means of adaptive optics was recently introduced by Bilsing~\citet{Bilsing2024a}.
%In this approach, first, the distortion by the oscillating gas-liquid interface is detected from reflected light by a wavefront sensor, and subsequently corrected in real-time through a deformable mirror.
This method, however, relies on a sophisticated optical system and has been tested only for a relatively flat droplet and low oscillation amplitude so far.
Conversely, the ray-tracing method for distortion correction can be expanded straightforwardly to arbitrary droplet shapes, if the instantaneous three-dimensional shape of the gas-liquid interface can be acquired.

%Volumetric interface reconstruction}
For the volumetric reconstruction of interfaces in two-phase flows, various methods based on different optical measurement techniques have been proposed.
Higashine~\citet{Higashine2008} reconstructed the 3D gas-liquid interface shape of droplets deformed by gravitational or centrifugal forces through an analytical model based on the three-dimensional Laplace equation.
% and an approximation of the droplets cross-section and contact line by circular and elliptical segments.
While the analytically calculated droplet shape agrees well with experiments, only the equilibrium state is considered and an accurate measurement of the advancing contact angle is required.
R{\'i}os-L{\'o}pez~\citet{RiosLopez2018} reconstructed a sliding droplet from orthogonal top and side views contours, recorded by the shadowgraphy technique.
%The reliance on the top view, however, introduces errors for droplets on hydrophobic substrates, due to a self-occlusion of interface.
Multi-camera techniques allow for a higher accuracy of the volumetric reconstruction \cite{Fu2018, Masuk2019}, but require more complicated experimental setups and unrestricted optical access.
Therefore, the scope of the present work is the application of a reconstruction technique based on a monocular shadowgraphy setup and deep learning that was previously introduced by the authors \cite{Dreisbach2023b}.
%for the spatio-temporal reconstruction of adhering droplets in shear flows.
The basis for the volumetric reconstruction are images obtained through a purposefully developed optical measurement technique, in which color-coded glare points are used in combination with the shadowgraphy technique in order to embed additional three-dimensional information of the gas-liquid interface in the images.
Subsequently, the droplet shape is reconstructed from the images by a neural network, which is trained for the physically correct reconstruction of the gas-liquid interface on numerical data.
The accuracy of the proposed technique is first evaluated by the reconstruction of synthetic image data and, subsequently, the robustness of the reconstruction is investigated in experiments with increasingly deformed oscillating droplets.

\section{Methodology}

A neural network that was developed for the volumetric reconstruction from monocular images \cite{Saito2019} is trained by supervised learning on a labeled dataset of droplet images and the respective three-dimensional gas-liquid interface shapes.
Numerical simulation provides suitable ground truth data of the gas-liquid interface that allows for the encoding of a physically correct model of the adhering droplet dynamics within the neural network.
However, the numerical simulation and the images obtained in the experiments do not perfectly agree due to differences in the initial conditions, uncertainty in the experiments, and errors from modeling and numerical approximations.
This matching problem between experiment and simulation is resolved by synthetic image rendering on the basis of gas-liquid interface shapes extracted from the results of the numerical simulation.
In that regard, the optical setup of the experiments is reproduced accurately in a rendering environment that allows for physically correct ray tracing.
Thereby, synthetic images can be generated that match the experimental recordings in visual appearance and have an exact substantive correspondence to the numerical ground truth.
%The rendering setup introduced by Dreisbach~\citet{Dreisbach2023b} is modified to reproduce the optical setup of the experiments and allow for the generation of synthetic image data that matches experimental recordings.
The neural network is trained for the physically correct reconstruction of the adhering droplet dynamics on the synthetic dataset and then employed for the reconstruction of real image data from the experiment.

In the following the fluid mechanical and optical setup of the experiments is presented in subsection~\ref{subsec:experimental_setup}, which is followed by the description of the training data generation in subsection~\ref{subsec:train_data}.
The methodology for the training of the neural network and the prediction from experimental recordings is detailed in subsection~\ref{subsec:vol_rec} and the metrics used for the evaluation of the reconstruction performance are introduced in subsection~\ref{subsec:eval_metric}.

\subsection{Experimental methods}
\label{subsec:experimental_setup}

\begin{figure}[htbp!]
    \centering
    \def\svgwidth{1.0\columnwidth}
    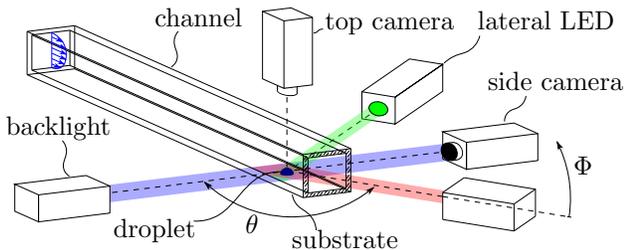
   \caption{Sketch of the measurement setup. Indicated are the scattering angle $\theta$ and the elevation angle $\Phi$ of the lateral light sources that determine the glare point behavior.}
   \label{fig:exp_setup-sketch}
\end{figure}

In order to enable the volumetric reconstruction of the adhering droplet's shape, an optical measurement technique that embeds additional three-dimensional information in the images is employed.
To that end, the previously introduced method by the authors \cite{Dreisbach2023}, which extends the canonical shadowgraphy technique by colored glare points from additional lateral light sources is employed.
The optical setup is adapted and integrated with a flow channel that allows for the investigation of adhering droplets in external shear flows at different fluid mechanical conditions.

\begin{figure}[htbp!]
\centering
    \includegraphics[width=\linewidth]{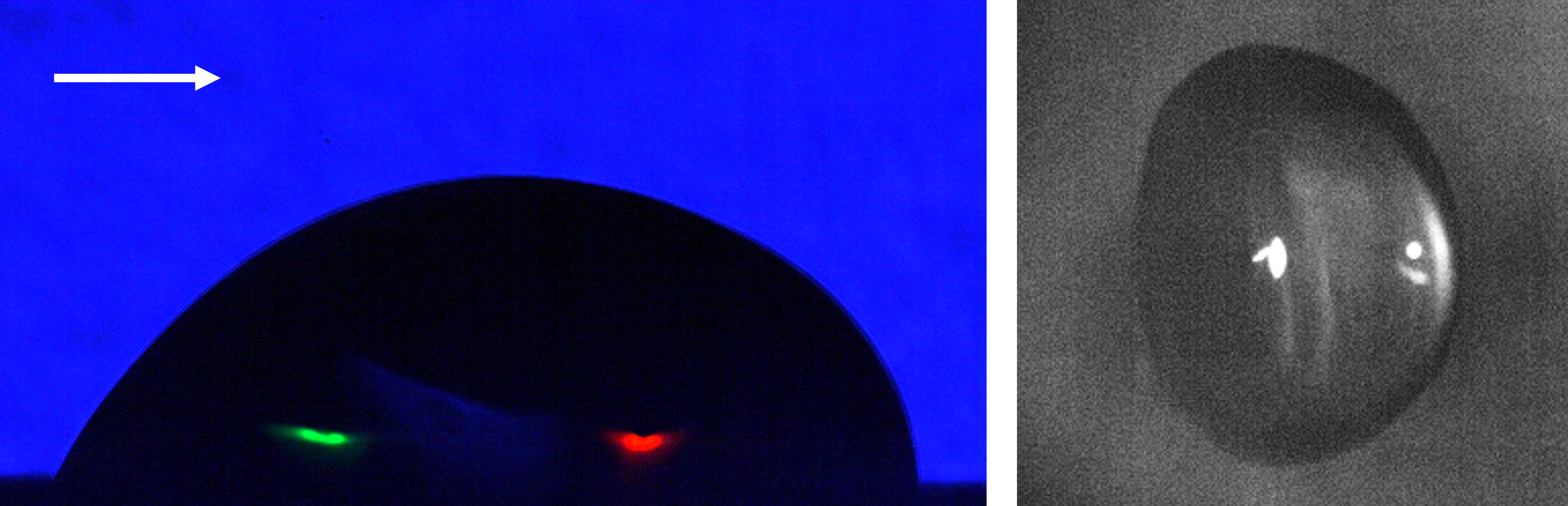}
    \caption{Raw images recorded by the side camera (left) and the top camera (right). In both images, the main flow direction is from left to right.}
    \label{fig:images_exp}
\end{figure}

Within the experimental arrangement, as shown in Figure~\ref{fig:exp_setup-sketch}, a blue LED light is used as backlight for the shadowgraphy setup, which produces an accurate projection of the droplet's contour in the image.
Additionally, two lateral red and green LED light sources are mounted at specific scattering and elevation angles with respect to the droplet in order to create colored glare points on the gas-liquid interface.
Three high-power \emph{ILA\_5150 LPSv3} LEDs with narrow-banded spectra and maxima in the visible spectrum at $\sim 455$\,nm ({blue}), $ \sim 521$\,nm ({green}) and $\sim 632$\,nm ({red}) are used as illumination sources.
Given the smoothness of the phase boundary, pure interface reflection can be considered for the lateral glare points.
An additional glare point resulting from the two-fold refraction of the backlight during entering and exiting the gas-liquid interface appears in the center of the droplet, as can be seen in Figure~\ref{fig:images_exp}.
As the geometric setup of the light propagation is known, additional three-dimensional information of the gas-liquid is encoded in the position and the shape of the glare points as elaborated in previous work by the authors \cite{Dreisbach2023b}.
The shadowgraph images of the adhering droplet with glare points are recorded by a $36$-bit color dynamic range \emph{Photron Nova R2} equipped with a \emph{Nikon AF Micro Nikkor} $2.8/105$ macro lens at $7,500$ frames per second (fps) and $1,280$\,px x $512$\,px resolution.
Moreover, a second orthogonal viewpoint from the top is recorded to obtain reference images for the evaluation of the out-of-plane accuracy of the reconstruction.
For that purpose, a $12$-bit monochrome dynamic range \emph{Photron Fastcam SA4} equipped with a \emph{Zeiss Milvus} $2/100 \mathrm{M}$ macro lens is used, which records at $3,000$ fps and $1,024$\,px x $1,024$\,px.
%Both cameras are aligned using the flow channel as a common point of reference and synchronized by the switch of the light.

A flow channel with a $22$\,mm x $22$\,mm cross-section and $1010$\,mm length, made of acrylic glass (PMMA), is used in order to provide full optical access.
The water droplets are placed at the downstream end of the channel, where a fully developed laminar or turbulent flow can be ensured.
% around 966 mm away from the honeycomb
%Droplets in the wall-near region with highest gradients in velocity -> highest shear forces that lead to droplet deformation and oscillation
The volume flow rate of the air flow is measured by a \emph{Testo} $6451$ compressed air meter and adjusted by a pressure valve.
In order to investigate the influence of different degrees of droplet motion, the bulk velocity of the air flow is varied in three discrete steps at $u_\mathrm{B}=5.85$\,m/s, $u_\mathrm{B}=7.58$\,m/s, and $u_\mathrm{B}=8.32$\,m/s, while the droplet volume is varied between $5$ and $22$ \textmu l.
The corresponding Reynolds numbers, with respect to the bulk velocity and the hydraulic diameter of the channel, lie in the range of $\mathit{Re}_\mathrm{ch}=8,500 - 12,100$.

The substrate on which the droplets are placed consists of PMMA, which is moderately hydrophilic with an advancing contact angle of $85^\circ$ and a receding contact angle of $44^\circ$, measured by the tilting method \cite{Maurer2016}.
Note that dynamic contact angles lower than $90^\circ$ ensure that the top view on the contact line is never self-obstructed by the droplet's gas-liquid interface and thus a continuous tracking of the wetted area is possible.
However, the low dynamic contact angles necessitate an adjustment of the lateral light sources to an elevation angle of $60^\circ$ in order to ensure the occurrence of glare points on the gas-liquid interface in all frames.

\subsection{Training dataset}
\label{subsec:train_data}

The data used to train the neural network for the reconstruction of the adhering droplet's interface is obtained from numerical simulations conducted with the customized solver \textit{hysteresisInterFoam} implemented in the open-source computational fluid dynamics (CFD) software OpenFOAM\textsuperscript{\textregistered}.
The solver uses the algebraic volume of fluid (VoF) method and is customized to allow for the pinning of the contact line by means of a contact angle adjustment \cite{Kramer2021}.
Surface tension is modeled as a body force by the continuum surface force (CSF) model \cite{Brackbill1992}.
The geometry used in the simulation model is a channel of rectangular cross-section.
At the channel walls, the no-slip boundary condition and at the outlet the total pressure boundary condition is used. 
For the lower wall, the contact angle boundary condition is applied using the modified contact line treatment mentioned before.
A block-structured grid with approximately $770,000$ cells is used.
Cell expansion is applied to refine the mesh close to the bottom wall. 
The same numerical schemes as in \cite{Kramer2021} are used in the current work.
In agreement with the experiments, the simulation features a $20$ \textmu l water droplet deposited on a PMMA substrate in a fully developed turbulent channel flow at $\mathit{Re}_\mathrm{ch}=8,500$ \cite{Kramer2021, Burgmann2022}.
In order to reduce the computational requirements, the fully developed turbulent velocity profile measured in the experiments by Barwari~\citet{Barwari2019} is defined at the inlet and laminar flow calculations are performed.
The average numerical time step is $\Delta t \approx 1$ms and a total of $1.1$\,s are simulated.
The droplet is discretized by approximately $20$ cells in the vertical direction and $17$ cells in the streamwise and spanwise direction.
The gas-liquid interface of the droplet is retrieved at each time step by extracting isosurfaces of the volume fraction at $\alpha=0.5$.
%spatial resolution of the droplet, 20 cells in the vertical direction and 17 cells in length/flow direction at t=0
% block structured grid with 766,656 cells
%Zeitdisk. - 1st Order acc. impl. Euler
%Räuml.disk. -2nd Order acc. linear schemes
% Zeitschritte dynamisch über Courant number<=0.5 eingestellt, liegen alle in der Ordnung 1E-5

In order to accurately simulate the glare point behavior smooth surface meshes with a sufficiently high resolution are required.
For that purpose, the resolution is increased by three consecutive iterations of mesh subdivision using linear interpolation \cite{Dyn2002}, followed by a smoothing of the mesh with the Taubin filter \cite{Taubin1995} for $50$ iterations.
%linear subdivision: Each subdivision creates 4 new triangles
%Mesh smoothing with Taubin filter - approximately conserves volume
In previous work, the authors introduced a rendering setup for the generation of synthetic images in the rendering environment \textit{Blender} \cite{Blender} with the \textit{LuxCore} \cite{LuxCore} package, which allows for physically correct ray tracing \cite{Dreisbach2023b}.
This rendering setup is adapted to accurately reproduce the optical setup of the experiments and is subsequently employed to generate synthetic glare-point shadowgraph images for each surface mesh.
%obtained from the numerical simulation
% The channel is explicitly not modeled, as refraction of light by the wall of the channel does not influence the position of the glare points, since the position is purely determined by the local angle of the gas-liquid interface. (For a large cone of light only a small glare point results in the interface at the position that matches the scattering angle, light cones larger than droplet, therefore translation of light by refraction has no influence on position, only a small influence on intensity, which can be corrected/modeled by light intensity directly)

\begin{figure}[htbp!]
\centering
    \includegraphics[width=\linewidth]{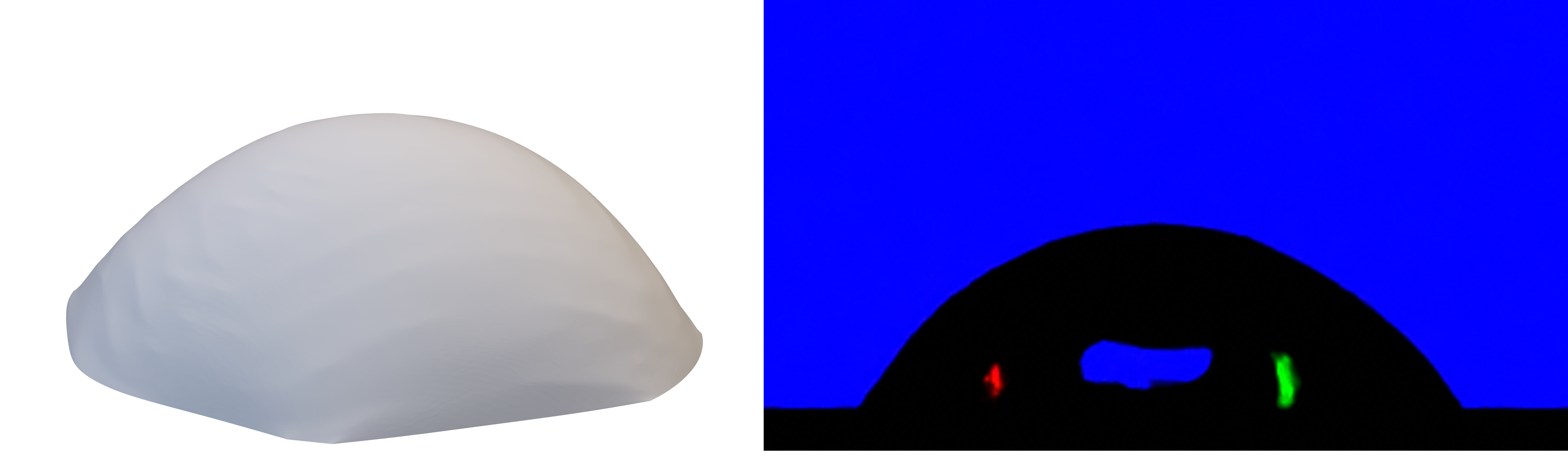}
    \caption{Ground truth gas-liquid interface extracted from numerical simulation (left) and corresponding synthetically rendered image (right).}
    \label{fig:image_render}
\end{figure}

An example of a synthetically rendered image and the corresponding ground truth droplet shape are shown in Figure~\ref{fig:image_render}.
During the rendering of each time step, the droplet is rotated around the vertical axis by $360^\circ$ in $10^\circ$ increments in order to generate a sufficiently large training dataset, which in turn ensures the generalisability of the neural network to novel shapes.
By that measure, a dataset of $1,134$ ground truth three-dimensional droplet shapes with respective $40,824$ rendered images is obtained.
The dataset is split by a ratio of $0.7/0.1/0.2$ into training and separate validation and testing subsets.

\subsection{Volumetric reconstruction}
\label{subsec:vol_rec}

The PIFu neural network \cite{Saito2019}, which was developed for the volumetric reconstruction from monocular images, is trained on the synthetically rendered dataset for the physically correct reconstruction of the gas-liquid interface.
The neural network learns an implicit representation of the spatio-temporal droplet geometry in the form of a level-set function.
The continuous nature of the level-set function allows for the reconstruction at an arbitrary resolution, independent of the training data resolution.
As shown by the successful application to impinging droplets \cite{Dreisbach2023b}, the encoded information in the glare points facilitates the inference of the out-of-plane component of the droplet shape.
The training images are pre-processed by the superimposition of a binary mask that covers the substrate ground in the images.
This introduction of prior knowledge simplifies the reconstruction task, as the differentiation between the shadowgraph contour of the droplet and the liquid-solid contact area does not need to be learned by the neural network.
In order to allow for the reconstruction of different droplet volumes, the input images and the corresponding three-dimensional ground truth shapes are scaled randomly, which is a common data augmentation technique to introduce scale invariance to neural networks \cite{Shorten2019}.
The PIFu neural network is trained by the RMSProp optimization algorithm \cite{Tieleman2012} for eight epochs with hyperparameters according to \cite{Dreisbach2023b}, in particular a batch size of $12$ and learning rate of $0.001$, which is reduced by a factor of ten at the beginning of epochs six and eight.

The images obtained in the experiments are pre-processed prior to the reconstruction in order to further enhance the similarity to the synthetic images, which is crucial for an optimal generalization of the neural network between both image data domains \cite{Csurka2017,Shrivastava2017}.
The mutual perturbation in the color channels of the RGB image that resulted from cross-talk in the camera sensor and the polychromatic light of the LEDs is compensated by the correction method introduced by Dreisbach~\citet{Dreisbach2023}. 
Thereby, the colored glare points and the shadowgraph contour are separated into the respective image channels for each light source color.
In the second step, the substrate ground is masked analogously to the training data.
Subsequently, each frame of the video sequence recorded by the lateral camera in the experiments is reconstructed by the neural network, in order to obtain the instantaneous three-dimensional gas-liquid interface shape at each time instance.

\subsection{Evaluation metrics}
\label{subsec:eval_metric}

The accuracy of the reconstruction is evaluated on the basis of the reconstructed volumetric droplet shapes and depending on the availability of ground truth data.
The following metrics are used for evaluation:

\begin{itemize}
    \item The three-dimensional intersection over union 
    \begin{equation}
    \mathrm{3D\texttt{-}IOU} = \frac{R \cap GT}{R \cup GT}\label{eq:3D-IOU}
    \end{equation}
    is calculated by the fraction of the intersection volume and the union volume of the reconstructed droplet $R$ and the ground truth $GT$.
    The 3D-IOU is an extension of the 2D-IOU \cite{Everingham2010} to three-dimensional space, and therefore serves as a measure for the spatial volumetric accuracy of the reconstruction.
    \item The measured uncertainty of the reconstructed volume 
    \begin{equation}
    \sigma_\mathrm{V} = \frac{1}{\overline{V}} \sqrt{\frac{1}{n-1}\sum_{i=1}^{n} (V_{\mathrm{R},i} - \overline{V})^2}.\label{eq:sigmaV}
    \end{equation}
    is calculated by the standard deviation \cite{Bendat2010} of the volume of the reconstructed droplets over the course of one experiment $V_{\mathrm{R},i}$, normalized by the arithmetic mean $\overline{V}$ of the reconstructed volumes.
\end{itemize}

\section{Results and Discussion}

First, the proposed reconstruction technique is evaluated on synthetic data in subsection~\ref{subsec:val_synth} in order to validate the capability of the method for learning the spatio-temporal dynamics of the adhering droplet and to determine the accuracy of the reconstruction.
This is followed by the investigation of the reconstruction quality in the experiments at different fluid mechanical conditions and the generalization capability to different degrees of droplet deformation in subsection~\ref{subsec:res_real}.

\subsection{Validation on synthetic image data}
\label{subsec:val_synth}

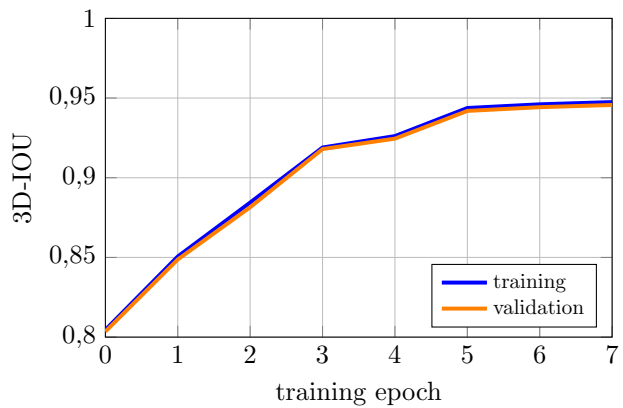
\begin{figure}[h]
    \centering
    \begin{tikzpicture}
        \begin{axis}[
        width=1.0\columnwidth,
        height=0.7\columnwidth,
        ymin=0.8,
        ymax=1,
        xmin=0,
        xmax=7,
        xlabel={training epoch},
        ylabel={3D-IOU},
        grid=major,
        legend cell align={left},
        legend pos=south east,
        legend style={nodes={scale=0.8, transform shape}},
        legend entries={training, validation},
        /pgf/number format/use comma,
        ]
        \addplot [line width=1.5pt, blue] table [x=epoch, y=train IOU, col sep=comma] {csv/DSH2024_train_val_IOU.csv};
        \addplot [line width=1.5pt, orange] table [x=epoch, y=val IOU, col sep=comma] {csv/DSH2024_train_val_IOU.csv};
        \end{axis}
    \end{tikzpicture}
    \caption{Average 3D-IOU of the reconstruction on the training and validation datasets during training of the network.}
    \label{fig:3D-IOU_train_val}
\end{figure}

The accuracy of the volumetric reconstruction was tracked during the training of the neural network through the evaluation of the average 3D-IOU calculated by the arithmetic mean over $1,000$ samples of the training and validation dataset at every epoch.
As can be seen in Figure~\ref{fig:3D-IOU_train_val} the neural network converges to a high volumetric accuracy on both the training and the validation dataset towards the end of the training ($\mathrm{3D{\texttt{-}}IOU}_\mathrm{train}=0.947$ and $\mathrm{3D{\texttt{-}}IOU}_\mathrm{val}=0.946$), which is close to a perfect agreement with the ground truth ($\mathrm{3D{\texttt{-}}IOU}_\mathrm{ideal}=1$).

The comparative analysis of all validation samples reveals that the accurate volumetric shape was reconstructed consistently throughout the dynamic deformation and oscillation of the droplet, as indicated by a low standard deviation of $\sigma_{\mathrm{3D{\texttt{-}}IOU}} = 0.027$.
The median of the volumetric accuracy on the validation dataset at the final training epoch was notably higher than the arithmetic mean ($\mathrm{3D{\texttt{-}}IOU}_\mathrm{median}=0.954$ and $\mathrm{3D{\texttt{-}}IOU}_\mathrm{val}=0.946$), due to outliers in the first five time steps of the validation dataset, which had a minimal accuracy of $\mathrm{3D{\texttt{-}}IOU}_\mathrm{min}=0.793$.
The shape of these outliers deviated significantly from the rest of the data distribution, thus explaining the lower reconstruction accuracy. %see Appendix
% train: 0.947460, val: 0.945657
% validation samples - std 0.02744, median: 0.9544
The overall high spatial accuracy of the prediction for the validation dataset, however, implies a good generalization to unseen samples.
These results indicate that the neural network successfully learned a spatio-temporal representation of the adhering droplet from the synthetic training data.

\begin{figure}[ht!]
\centering
    \begin{subfigure}[h]{1.0\linewidth}
        \caption{in-plane cross-section}
        \includegraphics[width=\linewidth]{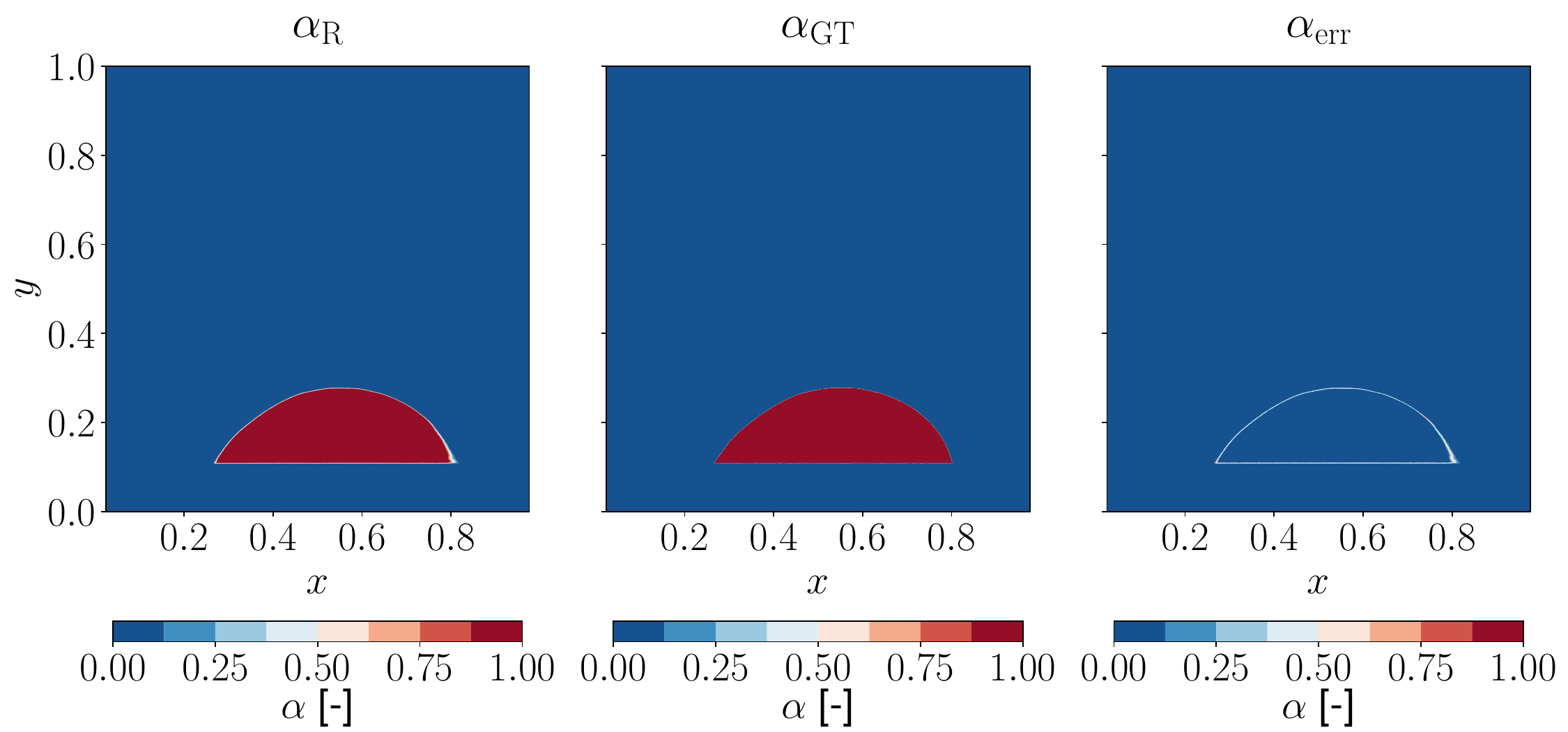}
        \label{fig:scs_first}
    \end{subfigure}
    \hfill
    \begin{subfigure}[h]{1.0\linewidth}
        \caption{out-of-plane cross-section}
        \includegraphics[width=\linewidth]{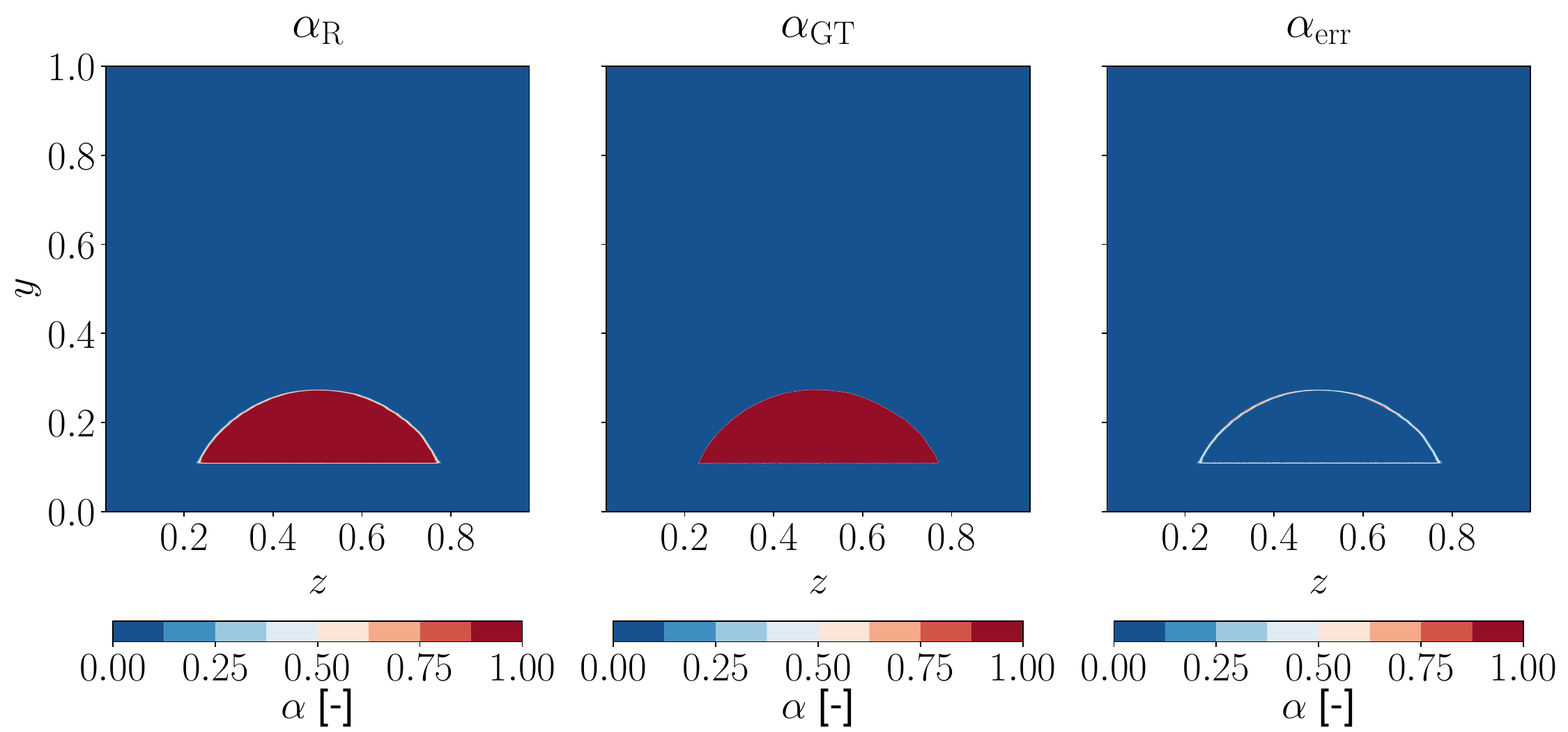}
        \label{fig:scs_second}
    \end{subfigure}
    \hfill
    \begin{subfigure}[h]{1.0\linewidth}
        \caption{wetted area} 
       \includegraphics[width=\linewidth]{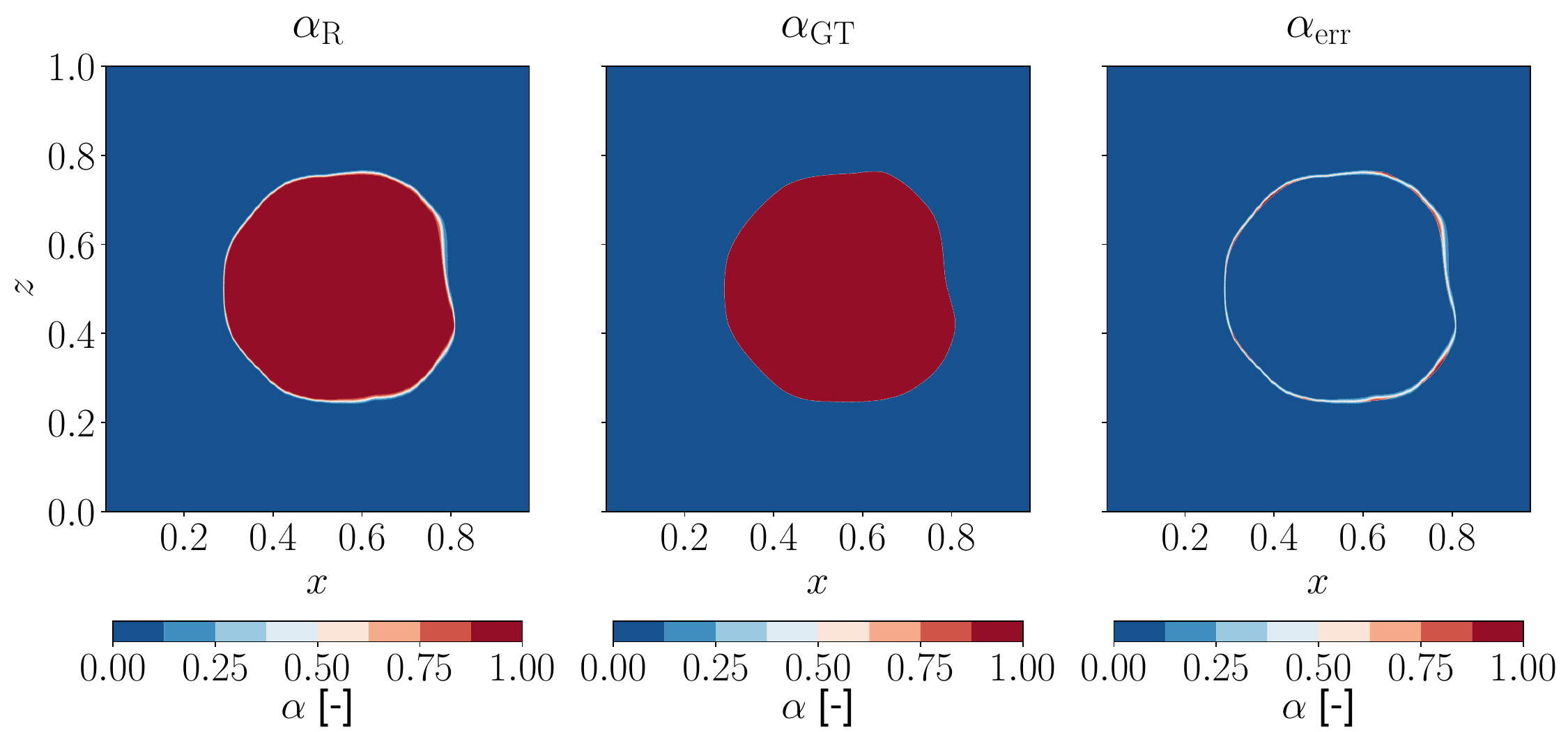}
        \label{fig:scs_third}
    \end{subfigure}
    \caption{cross-sections of the volume fraction $\alpha$; prediction by the neural network (left), ground truth (middle), and deviation of the prediction from the ground truth (right). In subfigures (a) and (c) the main flow direction is from left to right and in subfigure (b) the main flow direction is aligned to the image plane.
    \label{fig:synth_cross_sec}}
\end{figure}

Figure~\ref{fig:synth_cross_sec} shows the prediction of the phase distribution represented by the volume fraction $\alpha$ in three orthogonal cross-sections of the reconstructed volume.
The reconstruction results for a representative sample of the validation set are shown in comparison to the ground truth shape and the respective difference of both.
A volume fraction of one represents the liquid phase, whereas the gaseous phase has a volume fraction of zero, while any transitional values indicate the gas-liquid interface.

As can be seen, there is a good agreement with the ground truth in the global shape of the contour, as well as the local curvature of the reconstructed droplet shapes.
However, the gas-liquid interface of the reconstruction is more diffuse than the ground truth, which is a result of the continuous implicit representation of the droplet by the neural network.
In particular, the out-of-plane component of the reconstruction is more diffuse than the in-plane component, as can be seen by the comparison of Figures~\ref{fig:scs_first} and~\ref{fig:scs_second}.
Furthermore, the out-of-plane component of the droplet was found to be less accurately reconstructed, which was more obvious for high error samples (see Figure~\ref{fig:synth_cross_sec_high_error} in the Appendix). 
The more diffuse gas-liquid interface, as well as the higher deviation in the shape of the contour, indicate a higher uncertainty of the network's prediction in the out-of-plane direction.
This result was expected, as fewer image features, i.e. only the glare points, are available for the reconstruction of the out-of-plane component in comparison to the in-plane component, which has a strong basis for the reconstruction by the shadowgraph contour.
Therefore, the network has to rely more on the learned knowledge from the training data for the out-of-plane reconstruction.
These observations agree with the findings in previous work by the authors, in which the reconstruction of impinging droplets was considered \cite{Dreisbach2023b}.

As can be seen in Figure~\ref{fig:synth_cross_sec} the streamwise cross-section of the droplet is nearly planar-symmetrical, while the in-plane cross-section deviates significantly from planar symmetry due to the deformation of the droplet by the external flow.
Consequently, the in-plane geometry is more difficult to reconstruct, which is alleviated by the surplus of image features available for the reconstruction.
As demonstrated by the reconstruction results, both the planar symmetry of the streamwise cross-section and the deformed contour in the lateral view of the droplet are reconstructed accurately, which makes the proposed approach well-suited for the task of adhering droplet reconstruction.
% trades nicely with the setup of the experiments and resulting available information in image data for reconstruction
%Furthermore, the footprint of the droplet does not exactly adhere to planar symmetry, but reconstructed/shape inferred well by the network from images

\subsection{Reconstruction of experimental data}
\label{subsec:res_real}

\begin{figure*}[ht]
\centering
   %\includestandalone[width=1.0\textwidth]{Fig_envelopes}
   \includegraphics[width=1.0\textwidth]{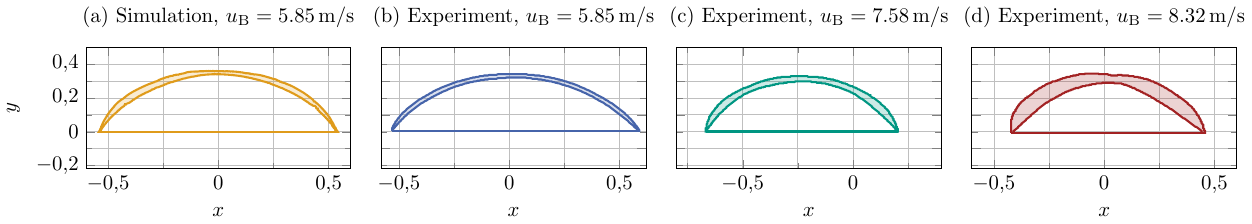}
    \caption{Envelopes of the droplet motion in the numerical simulation (left) and in the experiments (middle left to right) at different bulk velocities of external flow. Main flow direction from right to left.}
    \label{fig:envelopes}
\end{figure*}

The network trained on synthetic data was applied for the volumetric reconstruction from images obtained by experiments at different external flow velocities, in particular at the measured bulk velocities of $u_\mathrm{B}=5.85$\,m/s, $u_\mathrm{B}=7.58$\,m/s and $u_\mathrm{B}=8.32$\,m/s.
Figure~\ref{fig:envelopes} shows the envelopes of the droplet motion throughout the experiments for one representative case at each velocity in comparison to the numerical simulation.
As can be seen, an increase in the external flow velocity resulted in a larger degree of droplet deformation.
Furthermore, the comparison to the numerical data reveals that only at $u_\mathrm{B}=5.85$\,m/s a similar degree of deformation can be observed in the experiments.
%while external flow velocities of $u_\mathrm{B}=7.58$\,m/s and $u_\mathrm{B}=8.32$\,m/s lead to significantly higher degree of droplet motion.
Consequently, the neural network trained on the numerical data has to generalize to significantly more deformed and unknown droplet shapes in order to successfully reconstruct the experiments at $u_\mathrm{B}=7.58$\,m/s and $u_\mathrm{B}=8.32$\,m/s.
Note, that the same representative experiments (cases $1$, $8$, and $10$) will be the subject of the following evaluation unless stated otherwise.
The results of further experiments can be found in the appendix.

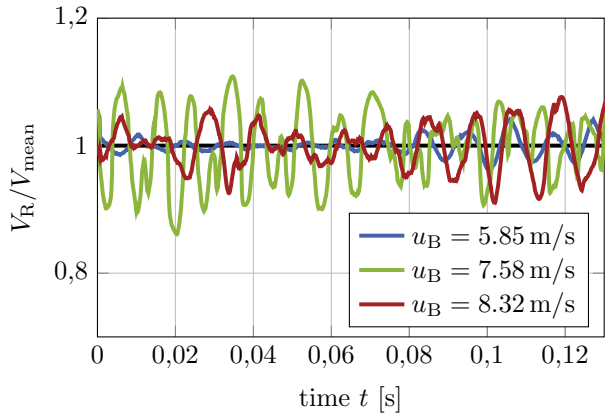
\begin{figure}[ht]
    \centering
    \begin{tikzpicture}
        \begin{axis}[
        width=1.0\columnwidth,
        height=0.7\columnwidth,
        ymin=0.7,
        ymax=1.2,
        xmin=0,
        xmax=0.13,
        xlabel={time $t$ [s]},
        ylabel={$ V_\mathrm{R}/V_\mathrm{mean}$},
        grid=major,
        legend cell align={left},
        legend pos=south east,
        legend style={nodes={scale=1.0, transform shape}},
        legend entries={$u_\mathrm{B}=5.85$\,m/s, $u_\mathrm{B}=7.58$\,m/s, $u_\mathrm{B}=8.32$\,m/s},
        /pgf/number format/use comma,
        scaled x ticks = false,
        x tick label style={/pgf/number format/fixed,
      /pgf/number format/1000 sep = \thinspace % Optional if you want to replace comma as the 1000 separator 
      }
        ]
        \addplot[color=black,line width=1.5pt,forget plot] coordinates {(-1,1) (1,1)};
        %%% 10.2
        \addplot [line width=1.5pt, KITblue] table [x expr=(\coordindex)/7500, y expr=\thisrowno{1}/0.15603039134044916, col sep=comma]{csv/Vol_and_Surf_Area_data_DSH2024_10.2_1.csv};
        %%% 13.2
        \addplot [line width=1.5pt, KITmaygreen] table [x expr=(\coordindex)/7500, y expr=\thisrowno{1}/0.11272646704932447, col sep=comma]{csv/Vol_and_Surf_Area_data_DSH2024_13.2_2.csv};
        %%% 14.5
        \addplot [line width=1.5pt, KITred] table [x expr=(\coordindex)/7500, y expr=\thisrowno{1}/0.12382670500974226, col sep=comma]{csv/Vol_and_Surf_Area_data_DSH2024_14.5_1.csv};
        \end{axis}
    \end{tikzpicture}
    \caption{Temporal evolution of the reconstructed droplet volume normalized by the mean volume for experiments with adhering droplets at different bulk velocities of external flow.}
    \label{fig:vol_comp}
\end{figure}

\begin{table}[ht]
    \centering
    \caption{Uncertainty $\sigma_\mathrm{V}$ of the reconstructed integral volume in percent of the mean volume for different external flow velocities averaged over all experiments conducted at a certain velocity.}
	\begin{tabular}{l|rrr}		%\toprule
            cases & $1-6$ & $7-9$ & $10,11$ \\ \hline \rule{0pt}{1.0\normalbaselineskip}$u_\mathrm{B}$ [m/s] & $5.85$ & $7.58$ & $8.32$ \\
            $\sigma_\mathrm{V} [\% \overline{V}]$ & $3.2$ & $4.1$ & $5.4$ \\
	\end{tabular} 
	\label{tab:uncert}
\end{table}

As no volumetric ground truth data is available for the reconstruction from images recorded in the experiments, the temporal evolution of the reconstructed droplet volume is considered for the evaluation of the reconstruction accuracy.
In the experiments no significant evaporation occurred within the time frame of one recorded image series ($1,000$ images over $133.3$\,ms) and, consequently, the volume of the droplet can be assumed to be constant.

Figure~\ref{fig:vol_comp} shows the temporal evolution of the normalized integral volume of the reconstructed droplet for one representative experiment at each evaluated velocity of the external flow.
As can be seen, the volume of the reconstruction fluctuates periodically around the mean, with an increasing amplitude towards higher velocities of the external flow, i.e. higher degrees of droplet deformation.
The measured uncertainties of the reconstruction, as detailed in Table~\ref{tab:uncert}, further underline that the reconstruction at higher degrees of droplet deformation is subject to higher uncertainty.
As indicated by the low uncertainty for the reconstruction at $u_\mathrm{B}=5.85$\,m/s the neural network trained on synthetic image data generalizes well to the reconstruction of experiments with a similar degree of droplet deformation to the numerical simulation that underlies the training data.
The successful reconstruction of the significantly more deformed gas-liquid interface in the experiments at $u_\mathrm{B}=7.58$\,m/s and $u_\mathrm{B}=8.32$\,m/s, which resulted in droplet shapes unknown to the neural network, indicate the trained network can extrapolate to different fluid mechanical conditions.
Furthermore, these results demonstrate the robustness of the method, which, however, is constrained by a reduced accuracy for the reconstruction of different data.

\begin{figure}[htbp!]
\centering
    \includegraphics[width=\linewidth]{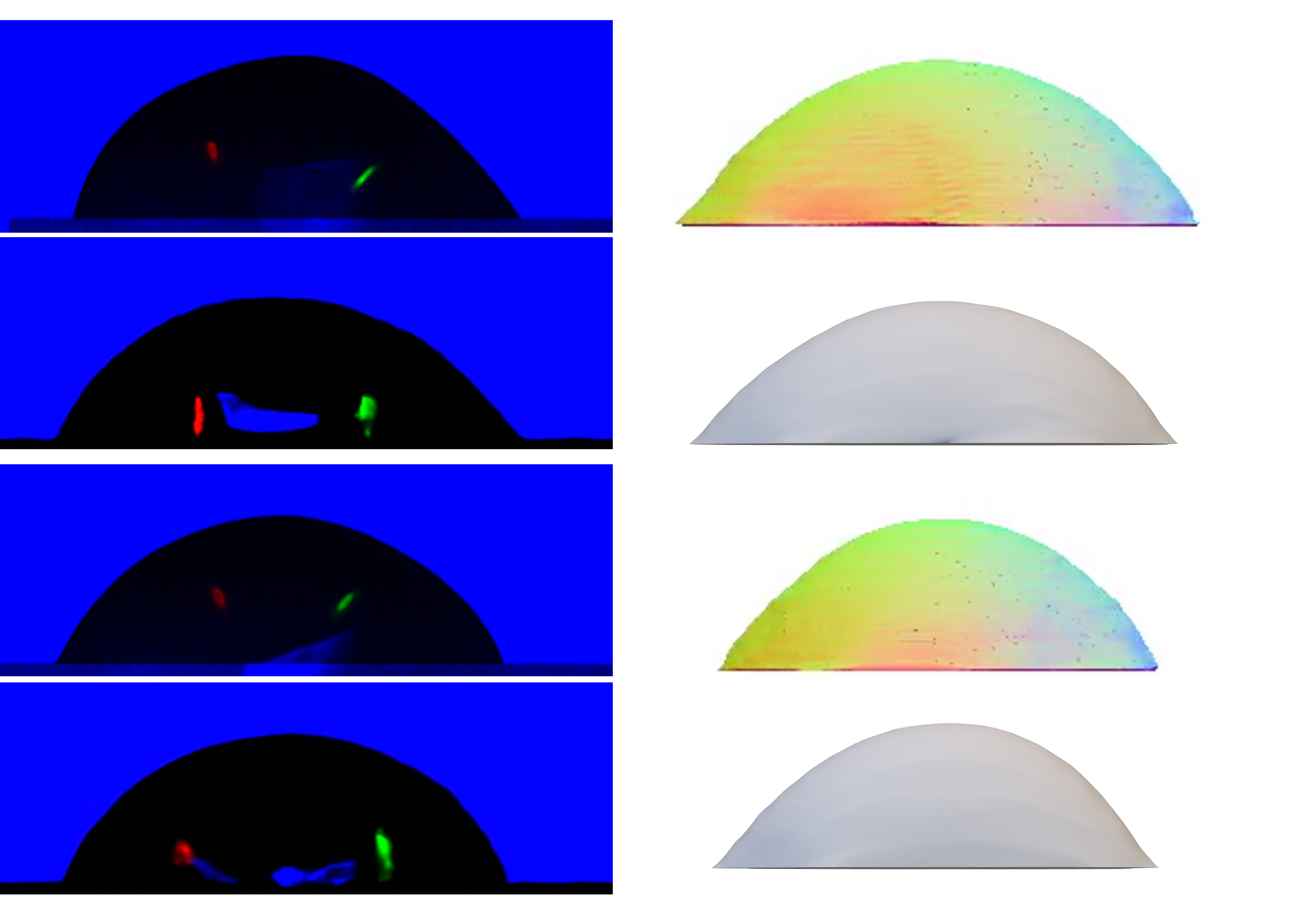}
    \caption{Input images (left) and out-of-plane projection (right) of the respective three-dimensional interfaces for the reconstruction at $u_\mathrm{B}=8.32$\,m/s. Rows one and three show the experimental recordings and the corresponding reconstructed interfaces and rows two and four show matching samples from the synthetic training dataset.}
    \label{fig:oscillation}
\end{figure}

% Side view
\begin{figure*}[!htb]
\centering
    %In-plane contours
    \begin{subfigure}[t]{0.325\textwidth}
        \caption{$u_\mathrm{B}=5.85$\,m/s}
       %\includestandalone[width=1.0\textwidth]{Fig_side_contour_10_2_1} 
       \includegraphics[width=1.0\textwidth]{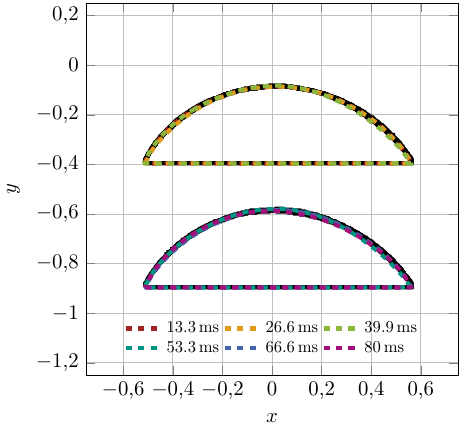}
        \label{fig:inpl_1}
    \end{subfigure}
    \hfill
    \begin{subfigure}[t]{0.325\textwidth}
        \caption{$u_\mathrm{B}=7.58$\,m/s}
        %\includestandalone[width=1.0\textwidth]{Fig_side_contour_13_2_2}
        \includegraphics[width=1.0\textwidth]{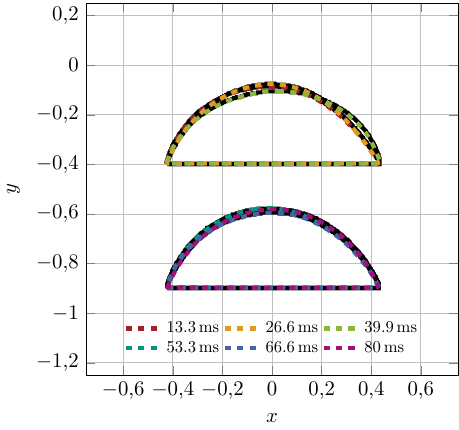}
    \label{fig:inpl_2}
    \end{subfigure}
    \hfill
    \begin{subfigure}[t]{0.325\textwidth}
        \caption{$u_\mathrm{B}=8.32$\,m/s}
        %\includestandalone[width=1.0\textwidth]{Fig_side_contour_14_5_1}
        \includegraphics[width=1.0\textwidth]{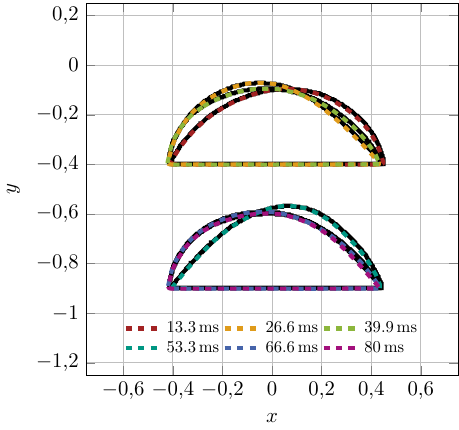}
    \label{fig:inpl_3}
    \end{subfigure}
    \caption{Temporal evolution of the reconstructed in-plane contour for different bulk velocities of external flow. The dashed colored lines indicate the reconstructed contours at different times and the black solid lines indicate the corresponding contours observed in the experiments. The main flow direction is from right to left.
    \label{fig:in-plane}}

    %Out-of-plane contours
    \hfill
    \begin{subfigure}[t]{0.325\textwidth}
        \caption{$u_\mathrm{B}=5.85$\,m/s}
       %\includestandalone[width=1.0\textwidth]{Fig_oop_contour_10_2_1}
       \includegraphics[width=1.0\textwidth]{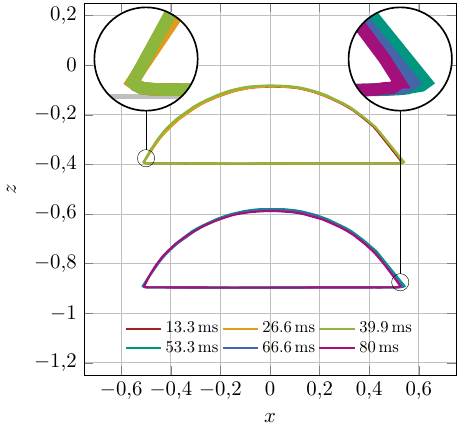}
        \label{fig:outpl_1}
    \end{subfigure}
    \hfill
    \begin{subfigure}[t]{0.325\textwidth}
        \caption{$u_\mathrm{B}=7.58$\,m/s}
        %\includestandalone[width=1.0\textwidth]{Fig_oop_contour_13_2_2}
        \includegraphics[width=1.0\textwidth]{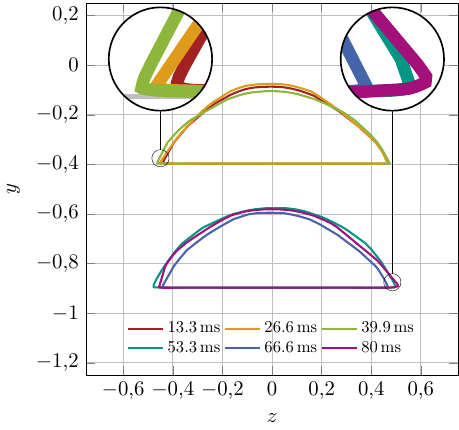}
    \label{fig:outpl_2}
    \end{subfigure}
    \hfill
    \begin{subfigure}[t]{0.325\textwidth}
        \caption{$u_\mathrm{B}=8.32$\,m/s}
        %\includestandalone[width=1.0\textwidth]{Fig_oop_contour_14_5_1}
        \includegraphics[width=1.0\textwidth]{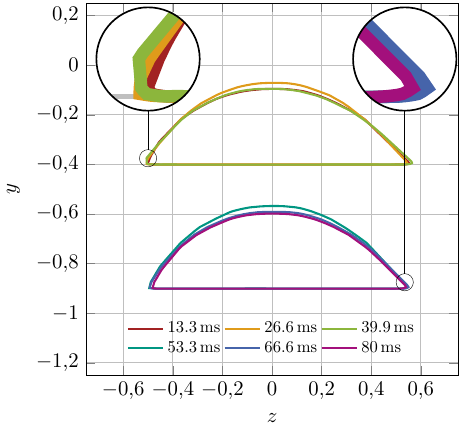}
    \label{fig:outpl_3}
    \end{subfigure}
    \caption{Temporal evolution of the reconstructed out-of-plane contour for different bulk velocities of external flow. The solid colored lines indicate the reconstructed contours at different times. The image plane is aligned with the main flow direction.
    \label{fig:out-plane}}
\end{figure*}

% Top view
\begin{figure*}[!htb]
\centering
    %Top view contours
    \begin{subfigure}[b]{0.325\textwidth}
        \caption{$u_\mathrm{B}=5.85$\,m/s}
        %\includestandalone[width=1.0\textwidth]{Fig_top_contour_10_2_1}
        \label{fig:wetted_area_a}
        \includegraphics[width=1.0\textwidth]{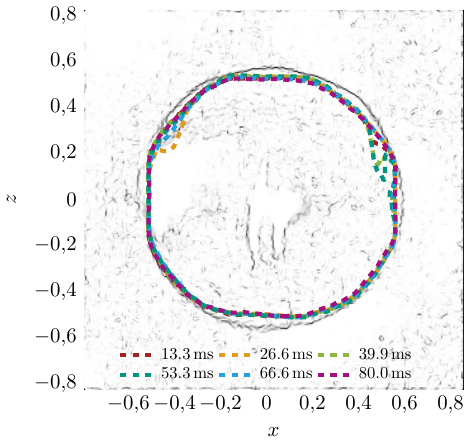}
    \end{subfigure}
    \hfill
    \begin{subfigure}[b]{0.325\textwidth}
        \caption{$u_\mathrm{B}=7.58$\,m/s}
        %\includestandalone[width=1.0\textwidth]{Fig_top_contour_13_2_2}
        \label{fig:wetted_area_b}
        \includegraphics[width=1.0\textwidth]{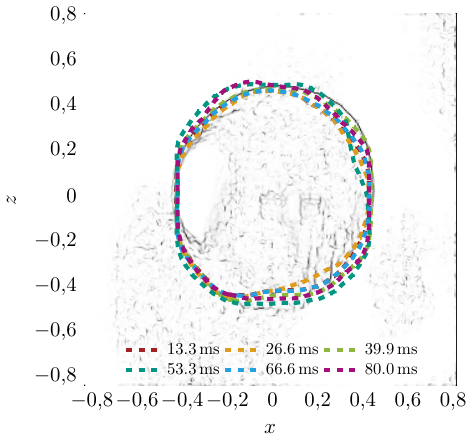}
    \end{subfigure}
    \hfill
    \begin{subfigure}[b]{0.325\textwidth}
        \caption{$u_\mathrm{B}=8.32$\,m/s}
       %\includestandalone[width=1.0\textwidth]{Fig_top_contour_14_5_1}
        \label{fig:wetted_area_c}
        \includegraphics[width=1.0\textwidth]{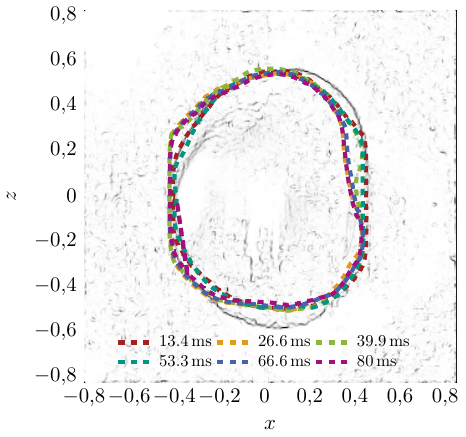}
    \end{subfigure}
    \caption{Temporal evolution of the reconstructed wetted area for different bulk velocities of external flow. The dashed colored lines indicate the reconstructed contours at different times and the grayscale image in the background indicates the corresponding contour observed in the experiments. The main flow direction is from right to left.
    \label{fig:wetted_area}}
\end{figure*}

%% Volume oscillation
The periodical oscillation of the reconstructed volume was found to coincide with the oscillation of the droplet contour.
Moreover, the input images to the volumetric reconstruction at the extrema of the volumetric deviation appear similar.
Figure~\ref{fig:oscillation} shows examples of the input images from the experiment that resulted in a local maximum (first row) and minimum (third row) of the volume and the respective out-of-plane projection of the reconstructed droplet shape.
In particular, the position of the glare points relative to each other and to the contour of the droplet was similar between the groups of input images that resulted in either the maxima or minima of reconstructed volume, which indicates a similar shape of the droplet interface.
Most saliently, a low position of the blue $p=1$ glare point in the input images resulted in a minimum volume, while a maximum in the reconstructed volume correlated with a high position of the $p=1$ glare point.

For both cases, synthetic images with similar relative glare point positions can be found in the training dataset, which are shown in rows two and four in Figure~\ref{fig:oscillation}.
Note, that Figure~\ref{fig:oscillation} shows sample images from the experiments at $u_\mathrm{B}=8.32$\,m/s for better visibility of the differences between the minimum and maximum case, as this experiment resulted in the highest deviations in the reconstructed volume.
Consequently, the contour of the droplet in the reconstruction is more deformed than the synthetic training data.
However, the reconstruction at lower velocities follows the same behavior.
The comparison of the out-of-plane projection between the reconstruction and the training data reveals a very similar shape of the contour.
%which indicates that the neural network predicts the depth according to the closest match of the input image to the training data distribution.
These results indicate that the neural network successfully learned the relation of the glare point positions to the three-dimensional geometry of the droplet and that this encoding of 3D information can be applied for depth estimation during reconstruction.
However, as the neural network has been trained for the reconstruction of different droplet volumes by an augmentation of the training data, as described in subsection~\ref{subsec:vol_rec} it is agnostic to the volume of the droplet.
Consequently, the shape of the contour is reconstructed with disregard to the integral volume of the droplet and thus the volume of the reconstruction is not conserved in time.

%% In-plane & Out-of-plane contours
In order to evaluate the in-plane and the out-of-plane accuracy of the reconstruction in more detail, two orthogonal projections of the reconstructed droplet interface are compared to the side- and top-view contours extracted from the corresponding images recorded in the experiments.
The results of the in-plane reconstruction for different external flow velocities are presented in Figure~\ref{fig:in-plane} and the results of the out-of-plane reconstruction are shown in Figure~\ref{fig:out-plane}.
As can be seen, there is an almost identical agreement of the reconstructed in-plane contours with the respective contours from the experiment, which underlines the findings of a high in-plane accuracy for the reconstruction of synthetic data.
The planar symmetry in the out-of-plane direction, which was expected due to the symmetry of the flow, was also reconstructed well for the experimental data.
%Furthermore, the interface in the out-of-plane direction is approximately smooth.
%As the neural network can be evaluated at any arbitrary resolution, high-quality meshes of the gas-liquid interface can therefore be reconstructed.
However, the reconstruction exhibits a fluctuation in the out-of-plane direction that is particularly noticeable in Figure~\ref{fig:outpl_2}.
This out-of-plane fluctuation is identified as the sole cause for the uncertainty of the volumetric reconstruction since the in-plane reconstruction is highly accurate. 
The out-of-plane reconstruction relies on the learned droplet geometry and the encoding of three-dimensional information by the glare points and, therefore, the accuracy of the reconstruction is reduced the more the input data deviates from the training data distribution.
The in-plane reconstruction is based on the image features, most importantly the contour of the shadowgraph, which provides a significantly larger amount and more direct information for the reconstruction.
As the in-plane reconstruction adheres closely to the contour in the images, the neural network adapts well to unknown shapes outside of the training data distribution.
These results fall in line with the observed uncertainty of the out-of-plane reconstruction for synthetic data in subsection~\ref{subsec:val_synth} and previous observations of impinging droplets \cite{Dreisbach2023b}.

% Top-view contours
The accuracy of the depth estimation is further evaluated by the comparison of the top-view projection of the reconstruction to the footprint of the droplet observed in the experiments through the top-view camera.
As can be seen in Figure~\ref{fig:envelopes}, the contact line of the droplet stays pinned during the duration of the experiments and the wetted area is therefore constant over time.
Figure~\ref{fig:wetted_area} shows the contour of the contact lines extracted from the experimental recordings at different external flow velocities.
The contours from the experiments are overlayed with the projected contours of the reconstructed droplet shapes for six different time instances.
As can be seen, the shape of the wetted area in the reconstruction is similar to the experiment and self-similar over time for $u_\mathrm{B}=5.85$\,m/s, while there is a larger deviation to the ground truth at higher external flow velocities.
The aspect ratio of the footprint changes significantly towards higher velocities as the droplet is deformed by the external flow.
Remarkably, the neural network successfully reconstructs these different aspect ratios, which is a significant extrapolation from the training data, which only contained one simulated case at the external flow velocity $u_\mathrm{B}=5.85$\,m/s, for which the footprint had an aspect ratio close to one.
For the same velocity in the experiments ($u_\mathrm{B}=5.85$\,m/s) a very good agreement of the reconstruction with the ground truth was reached.
At higher velocities ($u_\mathrm{B}=7.58$ and $u_\mathrm{B}=8.32$) the out-of-plane extent of the droplet was generally underestimated and reconstructed with a higher uncertainty, as indicated by the larger variation in the reconstructed footprints.
Furthermore, the unknown droplet shapes in the experiments related to higher aspect ratios of the wetted were reconstructed close to the distribution of shapes in the training data, thus revealing a bias of the trained model.
For all velocities the contour of the reconstruction is more angular in comparison to the experiment, but similar to the training data, as seen in Figure~\ref{fig:synth_cross_sec}, which is a further indication of model bias.
However, these results also demonstrate that the geometry of the droplet was learned faithfully to the training dataset by the network.
Therefore, the accuracy of the reconstruction is dependent on the truthful representation of the droplet dynamics in the training data.
The successful reconstruction of unknown droplet shapes with higher aspect ratios indicates that the network learned to utilize the glare points for the depth estimation.
%In addition to the relative position of the glare points, furthermore the defocus of the glare points changed with the depth of the droplet, thus providing additional cues for the depth estimation.

\subsection{Surface quality of the reconstructed interfaces}
%Resolution of the reconstruction

\begin{table}[ht]
    \centering
    \caption{Average time required for the reconstruction of one time step at different output resolutions.}
	\begin{tabular}{l|rrrr|r}		%\toprule
		grid nodes & $64^3$ & $128^3$ & $256^3$ & $512^3$ & $1,024^3$\\ \hline \rule{0pt}{1.0\normalbaselineskip}time [s] & 0.2 & 2.5 & 4.4 & 17.0 & 266.5 \\
	\end{tabular} 
	\label{tab:rec_time}
\end{table}

The implicit representation of the three-dimensional droplet geometry by the PIFu neural network \cite{Saito2019} allows for the reconstruction at an arbitrary resolution.
As higher resolutions are expected to result in a higher quality of the surface, but also increased computational costs, in the following the results of the reconstruction at different resolutions were evaluated.
Table~\ref{tab:rec_time} details the average time required for the reconstruction of one time step at different resolutions.
The domain for the reconstruction was discretized by an equidistant 3D grid with the same amount of grid nodes in all directions.
The trained PIFu network was sampled on this grid to predict the scalar occupancy field of the phase distribution.
Subsequently, the isosurface extraction algorithm marching cubes \cite{Lewiner2003,Lorensen1987} was employed to reconstruct the surface mesh from the occupancy field.
All calculations were performed on a single Nvidia RTX A5000 graphics processing unit.
As expected the computational costs grow quickly towards higher resolutions.
However, the octree \cite{Meagher1982} structure used during reconstruction significantly reduces the reconstruction time at higher resolutions up to $512^3$ grid nodes, which becomes obvious by the deviation from the cubic growth rule that would be expected otherwise.
Note that this trend only holds true until $512^3$ grid nodes due to given hardware limitations.
%At a resolution of $1,024^3$ grid nodes, the reconstruction time increased significantly due to memory limitations. 

\begin{figure}[htbp!]
\centering
    \includegraphics[width=\linewidth]{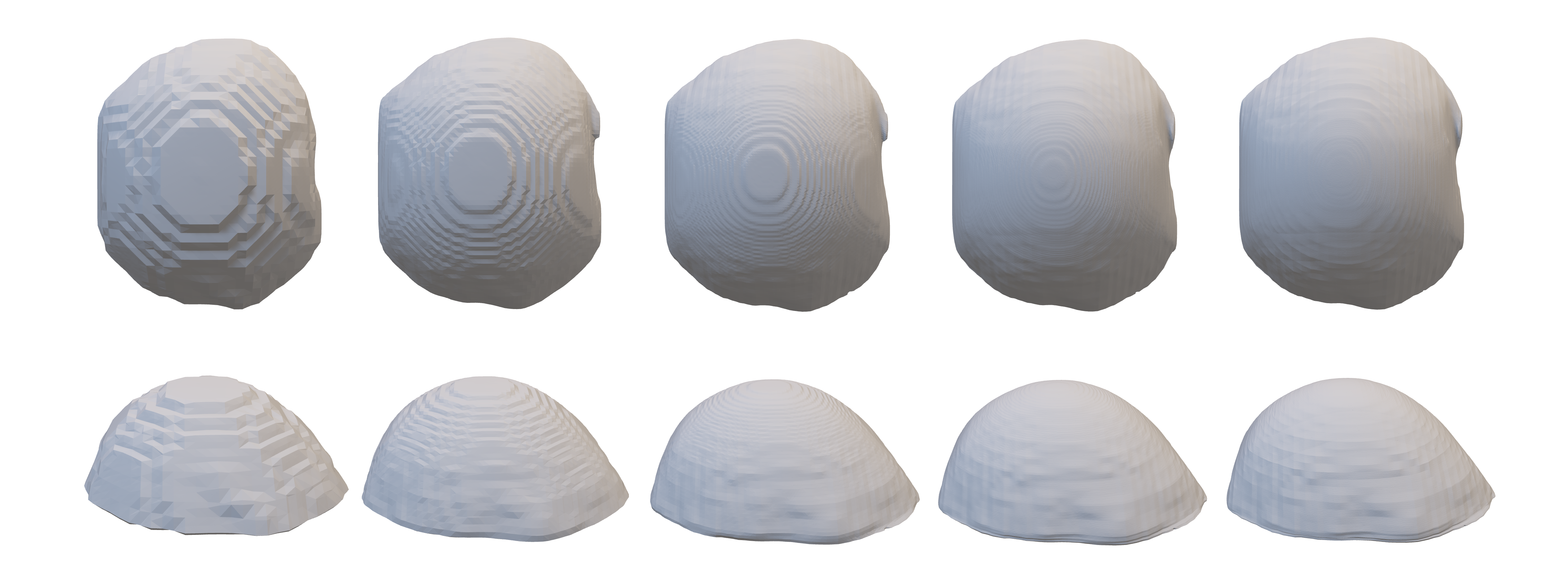}
    \caption{Volumetric reconstruction at different resolution for the same input image from the experiment at $u_\mathrm{B}=8.32$\,m/s, from left to right: $64^3$, $128^3$, $256^3$, $512^3$, $1024^3$ grid nodes resolution.}
    \label{fig:resolution}
\end{figure}

Figure~\ref{fig:resolution} shows the results of the volumetric reconstruction at different resolutions for the same input image obtained in the experiments at an external flow velocity $u_\mathrm{B}=8.32$\,m/s.
As can be seen, there is a significant visual improvement in the surface smoothness as the resolution increases from $64^3$ to $512^3$.
However, with the increase in the resolution from $512^3$ to $1,024^3$ grid nodes, the quality of the reconstruction did not increase further.
In fact, a marginal decrease of the surface quality can be observed, as small-scale ripples appear on the surface for a resolution of $1,024^3$ nodes, which are likely smoothed out by lower resolutions.
Considering the total time of $4.7$\,h required for the reconstruction of one experiment with $1,000$ frames, a resolution of $512^3$ grid nodes was found to be optimal within the scope of obtaining accurate and smooth surface meshes for distortion correction.
Note, that all reconstruction results presented in this work have a resolution of $512^3$ grid nodes.

It should be noted, that the presented reconstruction method is able to reconstruct the gas-liquid interface of the droplet at a significantly higher resolution compared to the training data.
The spatial resolution of the numerical simulation that underlies the training data was $20$ cells in the vertical direction and $17$ cells in the streamwise and spanwise direction at $t=0$.
These results demonstrate that the neural network has the capability to learn a highly accurate representation of the droplet geometry even from data much coarser than the targeted reconstruction resolution.
Furthermore, the approach of smoothing the ground truth gas-liquid interfaces that were extracted from the numerical simulation is validated by the positive reconstruction results.
In order to further enhance the surface quality of the reconstruction smoothing by means of the Taubin filter (see subsection~\ref{subsec:train_data}) is employed.

\begin{figure}[htbp!]
\centering
    \includegraphics[width=\linewidth]{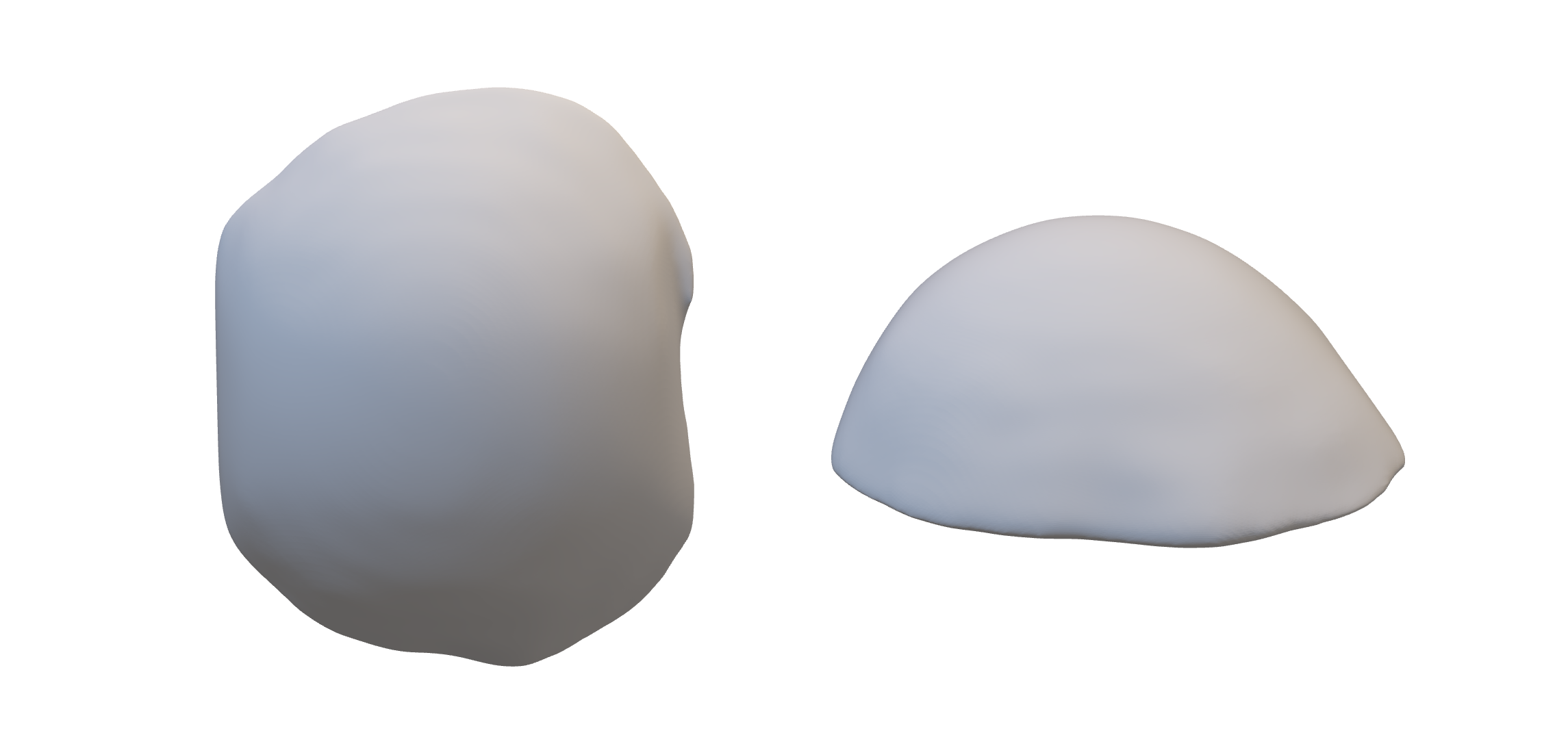}
    \caption{Smoothed volumetric reconstruction at $512^3$ grid nodes resolution.}
    \label{fig:rec_smoothed}
\end{figure}

As shown in Figure~\ref{fig:rec_smoothed}, the proposed process yields a smooth surface mesh of the droplet's gas-liquid interface that can be used for the intended distortion correction of PIV measurements.
Such distortion correction techniques require surface meshes that exhibit both a high fidelity to the true interface shape in the experiments and a smooth curvature, in order to allow for the accurate calculation of light refraction at the gas-liquid interface.
The prediction of the scalar occupancy field for the phase distribution by the PIFu neural network serves as an accurate basis for the surface reconstruction.
In this study, the efficient implementation \cite{Lewiner2003} of the marching cubes algorithm \cite{Lorensen1987} is employed to reconstruct the surface meshes.
However, alternative isosurface extraction algorithms, including extensions of the marching cubes algorithm \cite{Westermann1999, Kobbelt2001}, methods based on Delaunay triangulation \cite{Boissonnat2005}, or methods used in the numerical simulation of two-phase flows, such as the Piecewise-Linear Interface Calculation (PLIC) scheme \cite{Youngs1982}, may offer improved reconstruction performance.
Therefore, further research is required to assess the suitability of the reconstructed surfaces for distortion correction in the experiments.
This includes further development of surface reconstruction methods and smoothing operations, as well as experimentally validating the reconstructed meshes in the ray tracing approach to confirm their feasibility for distortion correction.

%As can be seen in Figure~\ref{fig:rec_smoothed} a smooth surface mesh can be obtained in this manner that allows for the accurate calculation of the refraction of light at the gas-liquid interface and thus is suitable for the intended distortion correction of PIV measurements.
%In order to be suitable for such distortion correction, the reconstructed surface mesh must exhibit both a high fidelity to the true interface shape in the experiments and a smooth curvature, so that the refraction of light at the gas-liquid interface can be calculated accurately.

\section{Conclusion}

The proposed method allows for the accurate reconstruction of the gas-liquid interface of adhering droplets in external shear flows from images recorded by means of a purposefully developed, yet simple optical measurement setup.
The evaluation using synthetic image data reveals that a high spatial accuracy can be reached by training the PIFu neural network on a synthetically rendered dataset based on numerical simulation.
Furthermore, the positive results for the reconstruction based on images recorded in the experiments demonstrate the applicability of the neural network trained on synthetic data to the real-world use case and thus validate the approach of training on synthetic data.

% In-plane vs out-of-plane -> learned geoemetry & depth encoding -> generalisation capability
In particular, the in-plane component of the reconstruction reaches an almost identical agreement to the ground truth, as the contour of the shadowgraph presents a strong basis for the in-plane reconstruction.
The out-of-plane reconstruction relies on the learned droplet geometry, in addition to the three-dimensional information encoded in the glare points, which is more sparsely distributed across the droplet interface.
Consequently, the uncertainty of the reconstruction is higher in the out-of-plane direction.
The successful reconstruction of adhering droplets at much higher flow velocities and greater deformation than in the training data shows that the method can effectively handle and predict new, unknown shapes.
It can be concluded that the combination of the learned geometry and the depth encoding by the glare points results in a robust and flexible model of the gas-liquid interface dynamics that can be used for the extrapolation to different flow conditions outside of the training data distribution.
The proposed method leverages the planar symmetry of the setup, simplifying the depth prediction along the streamwise direction.
Concurrently, the more significant deformation of the droplet in spanwise direction is imaged directly via the shadowgraph contour, thus reserving the previously mentioned high in-plane accuracy for the direction that undergoes more significant deformation.

% Suitability for distortion correction
The implicit representation of the interface by the neural network allows for the reconstruction at a fine resolution, while the training on the results of numerical simulation facilitates a high fidelity of the reconstructed contours to the underlying physics.
Therefore, the reconstructed gas-liquid interfaces are both spatially accurate and smooth, which makes the proposed method well-suited for distortion correction of PIV images, as the accurate reconstruction of the local curvature of the interface is important for the correct calculation of the refraction that causes the distorted velocity fields.
% Usefulness in comparison to current state-of-the-art
In comparison to the current state-of-the-art approach for distortion correction, in which rotational symmetry of the gas-liquid interface is assumed, the presented approach becomes increasingly beneficial towards higher external flow velocities that result in larger degrees of droplet deformation and thus non-axis-symmetrical droplet shapes.
%, as the neural network is able to predict the non-axis-symmetrical three-dimensional shape of the interface.

%Outlook
The addition of training data that represents the deformed interface of adhering droplets at higher external flow velocities is expected to increase the accuracy of the reconstruction.
Furthermore, the bias of the network towards certain geometries can most likely be reduced by a greater variation in the training dataset.
Thereby, the flexibility and robustness of the trained model could be further enhanced.
Further improvements in the reconstruction accuracy can be expected from an increased resolution of the numerical simulation that is used to source the training data. 
The effect is two-fold, on the one hand, more accurate three-dimensional ground truth data facilitates the learning of the droplet geometry and on the other hand, more finely resolved surface meshes would increase the quality of the synthetic images rendered by ray tracing.
%improvement of synthetic data generation towards a higher degree of realism in rendering
In future work, the introduction of physical constraints -- such as mass or momentum conservation -- into the training process via physics-informed neural networks \cite{Raissi2019} could be used to further increase the accuracy of the reconstruction.

\section{Author contributions}
Maximilian Dreisbach: Conceptualization, Writing - Original Draft, Investigation, Experiments,
Visualization, Software, Validation\\
Itzel Hinojos: Experiments, Writing - Review \& Editing\\
Jochen Kriegseis: Writing - Review \& Editing, Project administration, Resources\\
Alexander Stroh: Writing - Review \& Editing, Project administration, Resources\\
Sebastian Burgmann: Conceptualization, Writing - Review \& Editing, Specialist contact person for adhering droplet dynamics\\

\section{Acknowledgements}
The authors gratefully acknowledge the financial support from the Friedrich and Elisabeth Boysen Foundation (BOY-160).

\section{Data availability}
All data that support the findings of this study, including the weights of the trained neural networks and any supplementary files are uploaded to \href{https://doi.org/	10.35097/egqrfznmr9yp2s7f}{KITopen}.

\section{Appendix}

\begin{figure*}[htbp!]
\centering
   %\includestandalone[width=1.0\textwidth]{Fig_volume}
   \includegraphics[width=1.0\textwidth]{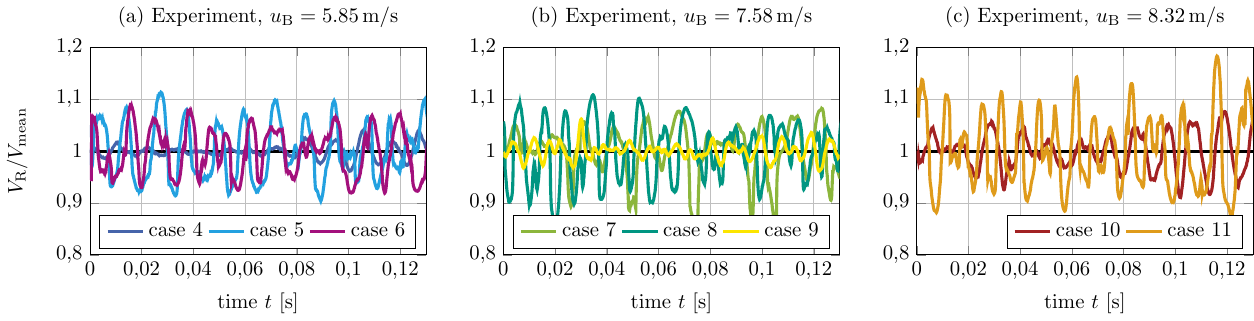} 
    \caption{Temporal evolution of the normalized integral volume of the reconstruction for adhering droplets in external shear flows at different bulk velocities $u_\mathrm{B}=5.85$\,m/s, $u_\mathrm{B}=7.58$\,m/s, and $u_\mathrm{B}=8.32$\,m/s.}
    \label{fig:app_vol_comp}
\end{figure*}

\begin{figure}[h]
\centering
    \begin{subfigure}[T]{1.0\linewidth}
        \caption{in-plane cross-section}
        \includegraphics[width=\linewidth]{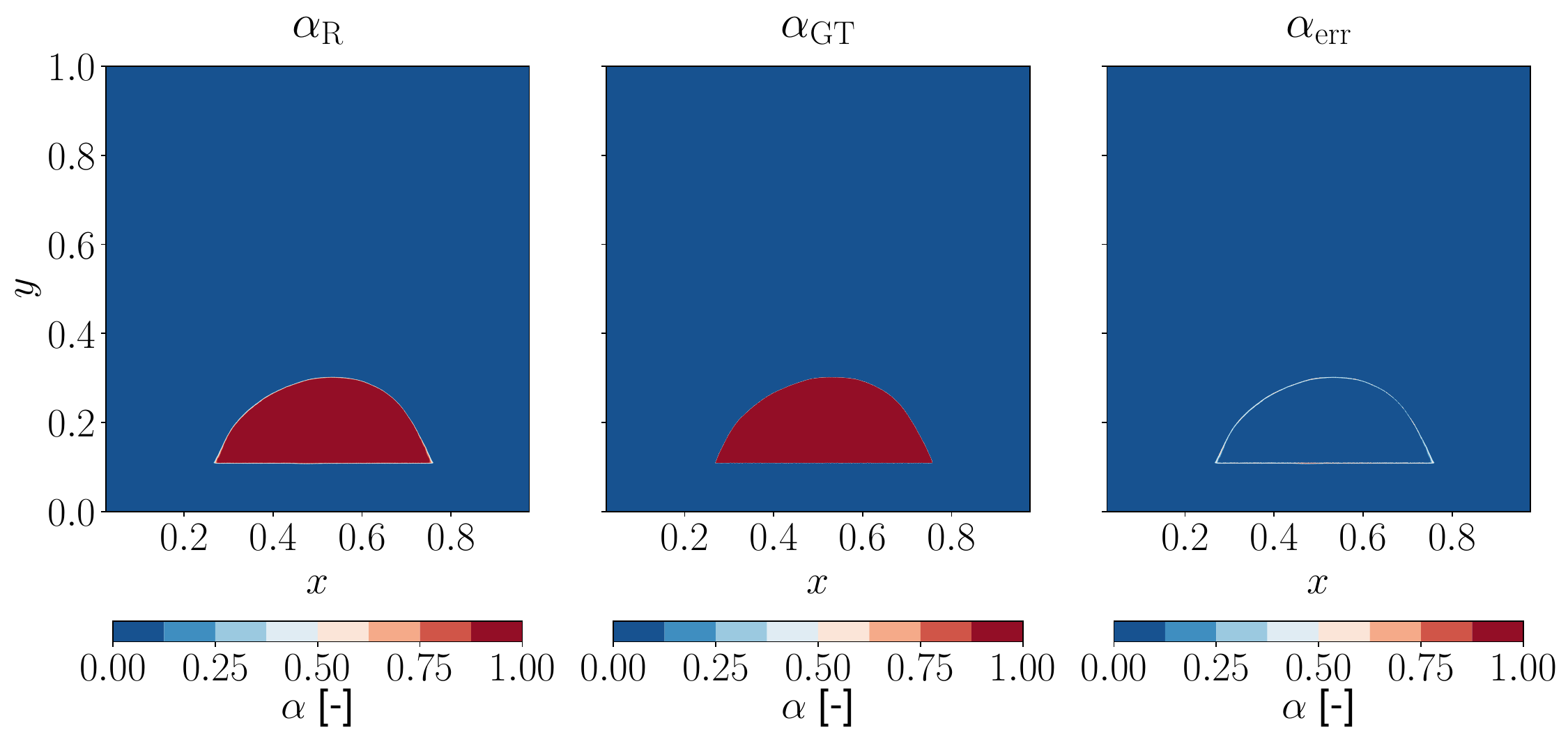}
        \label{fig:first}
    \end{subfigure}
    \hfill
    \begin{subfigure}[T]{1.0\linewidth}
        \caption{out-of-plane cross-section}
        \includegraphics[width=\linewidth]{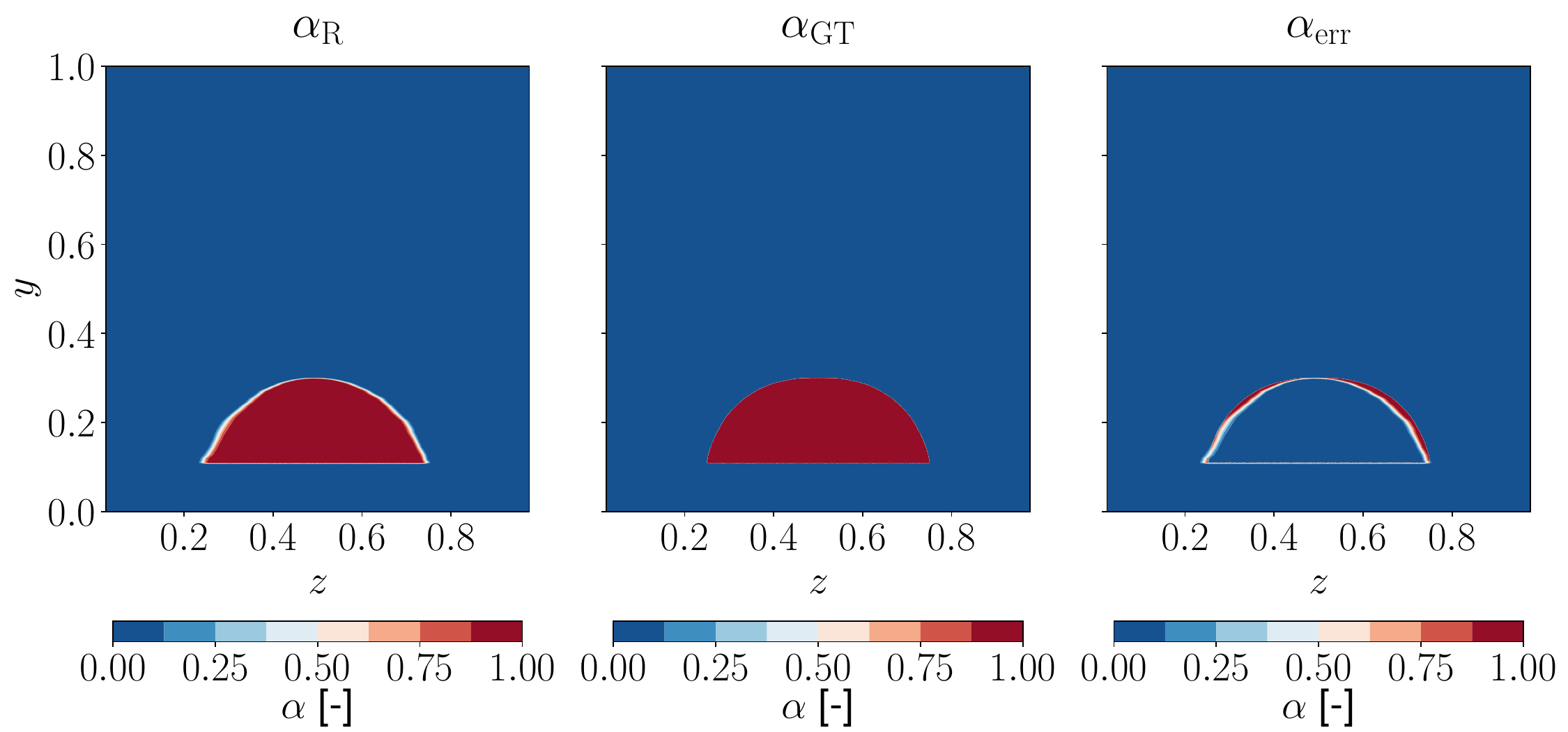}
        \label{fig:second}
    \end{subfigure}
    \hfill
    \begin{subfigure}[T]{1.0\linewidth}
        \caption{wetted area} 
       \includegraphics[width=\linewidth]{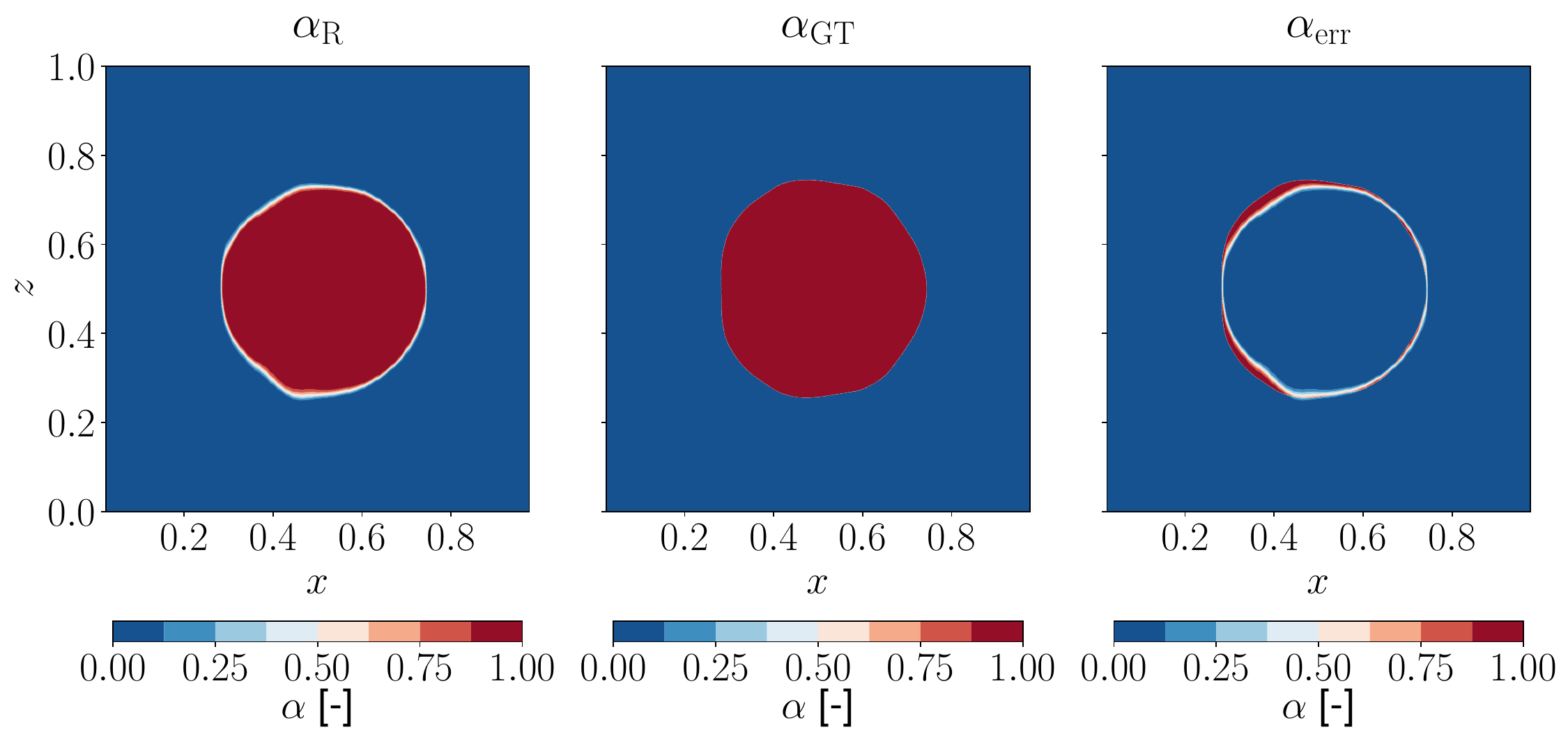}
        \label{fig:third}
    \end{subfigure}
    \caption{cross-sections of the volume fraction $\alpha$; prediction by the neural network (left), ground truth (middle), and deviation of the prediction from the ground truth (right). In subfigures (a) and (c) the main flow direction is from left to right and in subfigure (b) the main flow direction is aligned to the image plane.
    \label{fig:synth_cross_sec_high_error}}
\end{figure}

The temporal evolution of the normalized integral volume of the reconstructed droplet is presented in Figure~\ref{fig:vol_comp} for one representative experiment at each evaluated velocity $u_\mathrm{B}$ of the external flow.
These results are further detailed in Figure~\ref{fig:app_vol_comp}, showing the evolution of the reconstructed droplet volume for all cases at the respective velocities $u_\mathrm{B}=5.85$\,m/s, $u_\mathrm{B}=7.58$\,m/s, and $u_\mathrm{B}=8.32$\,m/s.
Figure~\ref{fig:synth_cross_sec_high_error} shows the reconstructed phase distribution represented by the volume fraction $\alpha$ in comparison to the ground truth data in three orthogonal cross-sections for one example of the validation data with high errors.
As can be seen, the out-of-plane reconstruction exhibits significantly larger errors in comparison to the in-plane reconstruction, which falls in line with the observations of the error distribution for a low error sample in Figure~\ref{fig:synth_cross_sec}.

\clearpage
\newpage
\section*{References}
\bibliographystyle{iopart-num}
%\bibliographystyle{unsrt}
% \providecommand{\url}[1]{{#1}}
% \providecommand{\urlprefix}{URL }
% \expandafter\ifx\csname urlstyle\endcsname\relax
%   \providecommand{\doi}[1]{DOI~\discretionary{}{}{}#1}\else
%   \providecommand{\doi}{DOI~\discretionary{}{}{}\begingroup
%   \urlstyle{rm}\Url}\fi
\bibliography{literature.bib}   % name your BibTeX data base

\end{document}

%% file: Drop_windtunnel.pdf_tex
%% Creator: Inkscape 1.1.2 (b8e25be833, 2022-02-05), www.inkscape.org
%% Accompanies image file 'Drop_windtunnel.pdf' (pdf, eps, ps)
\begingroup%
  \makeatletter%
  \providecommand\color[2][]{%
    \errmessage{(Inkscape) Color is used for the text in Inkscape, but the package 'color.sty' is not loaded}%
    \renewcommand\color[2][]{}%
  }%
  \providecommand\transparent[1]{%
    \errmessage{(Inkscape) Transparency is used (non-zero) for the text in Inkscape, but the package 'transparent.sty' is not loaded}%
    \renewcommand\transparent[1]{}%
  }%
  \providecommand\rotatebox[2]{#2}%
  \newcommand*\fsize{\dimexpr\f@size pt\relax}%
  \newcommand*\lineheight[1]{\fontsize{\fsize}{#1\fsize}\selectfont}%
  \ifx\svgwidth\undefined%
    \setlength{\unitlength}{1042.52581215bp}%
    \ifx\svgscale\undefined%
      \relax%
    \else%
      \setlength{\unitlength}{\unitlength * \real{\svgscale}}%
    \fi%
  \else%
    \setlength{\unitlength}{\svgwidth}%
  \fi%
  \global\let\svgwidth\undefined%
  \global\let\svgscale\undefined%
  \makeatother%
  \begin{picture}(1,0.40199604)%
    \lineheight{1}%
    \setlength\tabcolsep{0pt}%
    \put(0,0){\includegraphics[width=\unitlength,page=1]{Drop_windtunnel.pdf}}%
    \put(0.01894545,0.19152888){\makebox(0,0)[lt]{\lineheight{1.25}\smash{\begin{tabular}[t]{l}backlight\end{tabular}}}}%
    \put(0.77327577,0.3604202){\makebox(0,0)[lt]{\lineheight{1.25}\smash{\begin{tabular}[t]{l}lateral LED\end{tabular}}}}%
    \put(0.78941144,0.26204219){\makebox(0,0)[lt]{\lineheight{1.25}\smash{\begin{tabular}[t]{l}side camera\end{tabular}}}}%
    \put(0.52894562,0.36031806){\makebox(0,0)[lt]{\lineheight{1.25}\smash{\begin{tabular}[t]{l}top camera\end{tabular}}}}%
    \put(0.19072706,0.02760223){\makebox(0,0)[lt]{\lineheight{1.25}\smash{\begin{tabular}[t]{l}droplet\end{tabular}}}}%
    \put(0.2571361,0.36485933){\makebox(0,0)[lt]{\lineheight{1.25}\smash{\begin{tabular}[t]{l}channel\end{tabular}}}}%
    \put(0.47639802,0.0102377){\makebox(0,0)[lt]{\lineheight{1.25}\smash{\begin{tabular}[t]{l}substrate\end{tabular}}}}%
    \put(0,0){\includegraphics[width=\unitlength,page=2]{Drop_windtunnel.pdf}}%
    \put(0.40213463,0.03419053){\makebox(0,0)[lt]{\lineheight{1.25}\smash{\begin{tabular}[t]{l}$\theta$\end{tabular}}}}%
    \put(0.92628892,0.12687969){\makebox(0,0)[lt]{\lineheight{1.25}\smash{\begin{tabular}[t]{l}$\Phi$\end{tabular}}}}%
    \put(0,0){\includegraphics[width=\unitlength,page=3]{Drop_windtunnel.pdf}}%
  \end{picture}%
\endgroup%